\documentclass[aps, prb, twocolumn, 
superscriptaddress, amsmath, 
]{revtex4-2}

\usepackage{graphicx}
\usepackage{amsfonts}
\usepackage{amssymb}
\usepackage{amsmath}
\usepackage{lineno}
\usepackage{booktabs}
\usepackage{multirow}

\usepackage[normalem]{ulem}
\usepackage{hyperref}
\hypersetup{
    colorlinks=true,       
    linkcolor=blue,          
    citecolor=blue,        
    filecolor=blue,      
    urlcolor=blue           
}
\usepackage{makecell}
\usepackage{tabularx}
\usepackage{array}
\usepackage{float}

\newcommand{\ket}[1]{\vert#1\rangle}

\newcommand{\be}{\begin{equation}}
\newcommand{\ee}{\end{equation}}
\newcommand{\ba}{\begin{array}}
\newcommand{\ea}{\end{array}}
\newcommand{\bea}{\begin{eqnarray}}
\newcommand{\eea}{\end{eqnarray}}

\def\6{{\langle}}
\def\9{{\rangle}}

\usepackage{xcolor}

\begin{document}

\title{Experimental quantum telecloning across silicon photonic chips}

\author{Zicong Wen}
\thanks{These two authors contributed equally to this work}
\affiliation{National Laboratory of Solid State Microstructures, School of Physics, Collaborative Innovation Center of Advanced Microstructures, Nanjing University, Nanjing 210093, China}

\author{Kai Wang}
\thanks{These two authors contributed equally to this work}
\email{kai.wang@nju.edu.cn}
\affiliation{National Laboratory of Solid State Microstructures, School of Physics, Collaborative Innovation Center of Advanced Microstructures, Nanjing University, Nanjing 210093, China}

\author{Bochi Wu}
\affiliation{National Laboratory of Solid State Microstructures, School of Physics, Collaborative Innovation Center of Advanced Microstructures, Nanjing University, Nanjing 210093, China}

\author{Leizhen Chen}
\affiliation{National Laboratory of Solid State Microstructures, School of Physics, Collaborative Innovation Center of Advanced Microstructures, Nanjing University, Nanjing 210093, China}

\author{Yan-Qing Lu}
\affiliation{National Laboratory of Solid State Microstructures, School of Physics, Collaborative Innovation Center of Advanced Microstructures, Nanjing University, Nanjing 210093, China}

\author{Shining Zhu}
\affiliation{National Laboratory of Solid State Microstructures, School of Physics, Collaborative Innovation Center of Advanced Microstructures, Nanjing University, Nanjing 210093, China}
	
\author{Xiao-Song Ma}
\email{Xiaosong.Ma@nju.edu.cn}
\affiliation{National Laboratory of Solid State Microstructures, School of Physics, Collaborative Innovation Center of Advanced Microstructures, Nanjing University, Nanjing 210093, China}
\affiliation{Hefei National Laboratory, Hefei, 230088, China. }
\affiliation{Synergetic Innovation Center of Quantum Information and Quantum Physics, University of Science and Technology of China, Hefei, Anhui 230026, China.}

\date{\today}
\begin{abstract}
Telecloning—the combination of quantum teleportation and cloning—offers a powerful mechanism to disseminate unknown quantum states to multiple spatially separated recipients with optimal fidelity. Despite its conceptual importance for quantum networks, an experimental demonstration of symmetric qubit quantum telecloning remains elusive, particularly due to the challenges of generating multipartite entangled resource states and implementing stable multi-photon interference across distributed nodes. Here, we realize the optimal 1$\rightarrow$2 symmetric quantum telecloning using a scalable silicon photonic platform. We implement a six-photon protocol using two independent, fiber-linked photonic chips: one generating a heralded input state and the other preparing a four-photon entangled resource state $\ket{\Psi^-_4}$. By performing an interchip Bell-state measurement, we successfully distribute the input state into two optimal clones at remote nodes. We observe an interchip cloning fidelity of $78.45 \pm 1.39\%$, exceeding the classical limit of $2/3$ by 8 standard deviations. Our results demonstrate the robust generation and manipulation of complex multi-photon states between integrated chips, providing a foundational building block for large-scale multi-party quantum networks.
\end{abstract}

\maketitle

\section{Introduction}

The no-cloning theorem \cite{wootters1982single,DIEKS1982271} is a fundamental result of quantum mechanics which states that an unknown quantum state cannot be copied perfectly. This prohibition does not, however, preclude the construction of optimal approximate cloners that produce imperfect copies with the maximum fidelity permitted by quantum theory \cite{PhysRevA.54.1844,PhysRevLett.79.2153,PhysRevA.57.2368,PhysRevA.58.1827,RevModPhys.77.1225,Fan:2014tkq,PhysRevResearch.5.013139}. The theory of optimal quantum cloning — including both symmetric and asymmetric cloning — and the associated fidelity limits are well established \cite{RevModPhys.77.1225,PhysRevLett.79.2153,PhysRevA.57.2368,PhysRevA.58.1827,Cerf01022000,PhysRevA.72.042328,PhysRevLett.85.1754,PhysRevLett.80.4999,PhysRevLett.86.4938,PhysRevLett.86.4942,PhysRevA.109.022221}. Representative experimental demonstrations have been reported for both discrete-variable (DV) system \cite{Lamas-Linares:2002urx,PhysRevLett.92.067901,PhysRevLett.92.047902,PhysRevLett.95.030502,nagali2009optimal,PhysRevLett.106.180404,bouchard2017high}, and for continuous-variable (CV) system \cite{PhysRevLett.94.240503,PhysRevLett.98.170503,PhysRevLett.126.060503}. Cloning of quantum entanglement has also been investigated experimentally \cite{PhysRevLett.125.210502}. By contrast, quantum teleportation implements the faithful transfer of a quantum state from one location to another \cite{PhysRevLett.70.1895,bouwmeester1997experimental,PhysRevLett.80.1121} using pre-shared entanglement and classical communication.

Telecloning combines teleportation and cloning to distribute approximate copies of an unknown quantum state to multiple, spatially separated recipients \cite{PhysRevA.59.156}. Murao \textit{et al}. propose an efficient scheme, using multipartite entangled resource states together with local measurements and classical communication, to distribute optimal approximate clones of the input state to several distant parties \cite{PhysRevA.59.156}. 
Whereas distributing M identical qubits by independent quantum teleportation consumes M ebits and requires M separate teleportation events, the symmetric 
1 $\rightarrow$ M telecloning scheme \cite{PhysRevA.59.156} requires only $O(\log M)$ ebits of entanglement across the bipartition between the sender and the receivers. Here an ``ebit'' denotes the entanglement carried by a maximally entangled two-qubit (Bell) pair. For pure multipartite resource states we quantify the entanglement cost by the 
sender's reduced von Neumann entropy: $S(\rho_\mathrm{sender})=-\mathrm{Tr}[\rho_\mathrm{sender}\log_{2}\rho_\mathrm{sender}]$. This change in scaling — from $O(M)$ to $O(\log M)$ — relies on the use of multipartite entangled resource states. As required by the no-cloning theorem, there exists a fundamental trade-off between the number of output copies and their achievable fidelities \cite{RevModPhys.77.1225,gordon2007generalized}. Variants of telecloning include symmetric \cite{PhysRevA.59.156} and asymmetric \cite{PhysRevLett.95.030502} schemes, and CV telecloning \cite{PhysRevLett.87.247901,PhysRevLett.96.060504,PhysRevA.104.032419,PhysRevLett.132.160803}. Telecloning is of practical interest for quantum networks and distributed quantum information processing because it provides a way to disseminate quantum information to many nodes \cite{Fan:2014tkq,PhysRevA.72.032312}.


Practical telecloning requires simultaneous generation, coherent manipulation and low-loss distribution of multiple photons — demands that rapidly overwhelm bulk-optics approaches. Specifically, the realization of protocol of Murau \textit{et al.} is hampered by the difficulty of preparing the required resource state, which involves expanding a Bell pair into a class of special multiphoton entangled state. Moreover, the protocol relies on teleportation, resulting in a six-photon experiment. Consequently, despite the original scheme dating back to the 1999, its experimental demonstration has yet to be realized. Integrated quantum photonics mitigates these challenges by systematic integration of entangled photon sources, interferometric networks and routing with excellent stability, enabling higher phase stability, essential for multiphoton interference \cite{metcalf2014quantum,wang2018multidimensional,Zhang:2019dko,llewellyn2020chip,Lu:2020fqw,PhysRevLett.130.223601,zheng2023multichip,PhysRevLett.132.150602,Hoch2024Modular,Yin2025QuantumEnhanced,d53g-v8q6,Liu2025ChipToChip}. 

\begin{figure*}
\includegraphics[width=16cm]{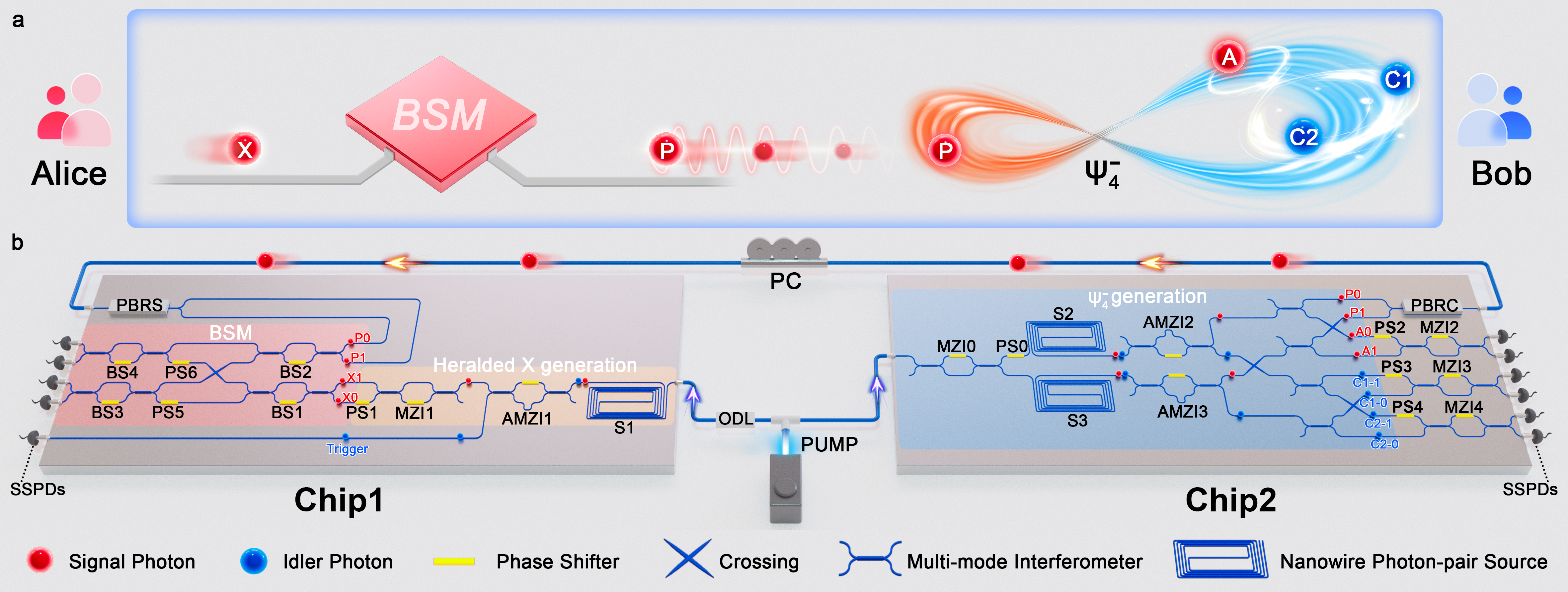}
\caption{Optimal symmetric quantum telecloning. \textbf{a}, Schematic of the 1$\rightarrow$2 telecloning protocol: Alice holds an unknown input photon X. 
Bob prepares the four-photon entangled state $\ket{\Psi^-_4}$ and sends the photon P to Alice. Alice then performs BSM on photon X and photon P from Bob.  Based on the BSM outcome of Alice, the input state of X is thereby cloned to photons C\textsubscript{1} and C\textsubscript{2}. \textbf{b}, Photonic-chip layout used to realize the telecloning. On Chip 1, source S\textsubscript{1} generates the trigger and the input photon X; the target state $\ket{\varphi}_\mathrm{X}$ is prepared by MZI\textsubscript{1} and PS$\textsubscript{1}$ and then interferes with the photon P (delivered from Chip 2) in the on-chip BSM module. On Chip 2, photon sources S\textsubscript{2} and S\textsubscript{3}, together with AMZIs, splitters and a routing network, produce — after post-selection on four-photon coincidences — the four-photon resource $\ket{\Psi^-_4}$ for photons P, A, C\textsubscript{1} and C\textsubscript{2}. The photon P is transmitted to Chip 1 via optical fiber. Conversion between the on-chip path encoding and the in-fiber polarization encoding is effected by PBRC/PBRS interfaces (see main text for details). MZI\textsubscript{2}-MZI\textsubscript{4} implement the single-qubit tomography for photons A, C\textsubscript{1} and C\textsubscript{2}.}
\end{figure*}

Leveraging these advantages, we report, to our knowledge, the first experimental realization of symmetric DV quantum telecloning. The photons are generated at about 1.5 µm wavelength, which benefits from low transmission loss in standard telecommunications fiber. We generate six-photon quantum state across two separate photonic chips. Chip 1 produces a heralded single photon X from a photon-pair source, which serves as the input state to be telecloned. On Chip 2, we generate the four-photon entangled state $\ket{\Psi^-_4}$ \cite{PhysRevA.59.156,PhysRevLett.101.010503} that serves as the telecloning resource for optimal symmetric 1 $\rightarrow$ 2 quantum telecloning. One photon P from this resource is routed to Chip 1. A Bell-state measurement (BSM) is performed on Chip 1 between photons X and P. Conditioned on a successful BSM, the quantum state of photon X is optimally cloned to photons C\textsubscript{1} and C\textsubscript{2} on Chip 2.

\section{Theory}

Below we introduce the 1→2 symmetric optimal quantum telecloning protocol \cite{PhysRevA.59.156} employed in this work (see Fig. 1a). Alice initially holds an unknown qubit X with its quantum state
\begin{equation}
|\varphi \rangle _\mathrm{X}=a|0\rangle_\mathrm{X}+b|1\rangle_\mathrm{X},\,\,    |a|^2+|b|^2=1,
\end{equation}
which she wishes to distribute two optimal clones of qubit X to remote parties (Bob) with equal fidelities. The protocol \cite{PhysRevA.59.156} requires a four-qubit entangled state $\ket{\Psi^-_4}$ prepared by Bob,
\begin{equation}
\label{eq:Psi4}
|\Psi _{4}^{-}\rangle _{\mathrm{PAC_1C_2}}=\frac{1}{\sqrt{2}}(|0\rangle _P\otimes |\phi _0\rangle _{\mathrm{AC_1C_2}}+|1\rangle _P\otimes |\phi _1\rangle _\mathrm{AC_1C_2})
\end{equation}
with

\begin{align}
|\phi _0\rangle _{\mathrm{A}\mathrm{C}_1\mathrm{C}_2}
=&\frac{1}{\sqrt{6}}
\Big[
|1\rangle _{\mathrm{A}}
\big(
|01\rangle _{\mathrm{C}_1\mathrm{C}_2}
+|10\rangle _{\mathrm{C}_1\mathrm{C}_2}
\big)
\nonumber\\
&\qquad
-2|0\rangle _{\mathrm{A}}
|11\rangle _{\mathrm{C}_1\mathrm{C}_2}
\Big],\\
|\phi _1\rangle _{\mathrm{A}\mathrm{C}_1\mathrm{C}_2}
=&\frac{1}{\sqrt{6}}
\Big[
|0\rangle _{\mathrm{A}}
\big(
|01\rangle _{\mathrm{C}_1\mathrm{C}_2}
+|10\rangle _{\mathrm{C}_1\mathrm{C}_2}
\big)
\nonumber\\
&\qquad
-2|1\rangle _{\mathrm{A}}
|00\rangle _{\mathrm{C}_1\mathrm{C}_2}
\Big],
\end{align}

where P denotes the qubit sent from Bob to Alice, A is an ancillary qubit retained by Bob, and $\mathrm{C}_1$, $\mathrm{C}_2$ are the two qubits that will be the clones. This resource state can be regarded as a Bell-type entangled state between photon P and the composite subsystem formed by photons A, C\textsubscript{1} and C\textsubscript{2}. Moreover, inside that composite subsystem, photon A is Bell-like entangled with the joint system C\textsubscript{1}, C\textsubscript{2}. Hence the overall resource closely relates to the concept of entangled entanglement \cite{PhysRevA.54.1793,PhysRevLett.97.020501,PhysRevA.106.053715}. Alice performs a Bell-state measurement on qubits X and P. The joint state of photons X, P, A, C\textsubscript{1} and C\textsubscript{2} are expanded with respect to the Bell state basis of photons X and P, {$\ket{\Phi^\pm}_\mathrm{XP}$, $\ket{\Psi^\pm}_\mathrm{XP}$}, yielding:
\begin{align}
\nonumber |\varphi \rangle _\mathrm{X}|\Psi _{4}^{-}\rangle _\mathrm{PAC_1C_2}&=\frac{1}{2}[|\Phi ^+\rangle _\mathrm{XP}(a|\phi _0\rangle _\mathrm{AC_1C_2}+b|\phi _1\rangle _\mathrm{AC_1C_2})\\
\nonumber &+|\Phi ^-\rangle _\mathrm{XP}(a|\phi _0\rangle _\mathrm{AC_1C_2}-b|\phi _1\rangle _\mathrm{AC_1C_2})\\
\nonumber &+|\Psi ^+\rangle _\mathrm{XP}(b|\phi _0\rangle _\mathrm{AC_1C_2}+a|\phi _1\rangle _\mathrm{AC_1C_2})\\
&-|\Psi ^-\rangle _\mathrm{XP}(b|\phi _0\rangle _\mathrm{AC_1C_2}-a|\phi _1\rangle _\mathrm{AC_1C_2})]
\end{align}

Conditioned on Alice's measurement outcome — for example $|\Phi ^+\rangle _\mathrm{XP}$, the state held by Bob becomes
\begin{equation}
|\Phi \rangle _\mathrm{AC_1C_2}=a|\phi _0\rangle _\mathrm{AC_1C_2}+b|\phi _1\rangle _\mathrm{AC_1C_2},
\end{equation}
which encodes the input qubit across Bob's three local qubits A, $\mathrm{C}_1$ and $\mathrm{C}_2$. To compensate for the Pauli rotations induced by the BSM, Bob applies a corresponding unitary correction (specifically, $\sigma_y$ for the $\ket{\Phi^+}$ outcome) to qubits $\mathrm{C}_1$ and $\mathrm{C}_2$.

Each clone is obtained by taking the reduced state of either $\mathrm{C}_1$ or $\mathrm{C}_2$. Tracing out the other subsystems yields the single-qubit density matrix:
\begin{equation}
\rho _{\mathrm{C}_i}=\frac{5}{6}|\varphi \rangle \langle \varphi |+\frac{1}{6}|\varphi ^{\bot}\rangle \langle \varphi ^{\bot}|,   i=1, 2
\end{equation}
so that the single-clone fidelity limit is 5/6.\\

The fidelity limit is independent of the input state and saturates the optimal value for symmetric 1→2 universal cloning of a qubit. This protocol implements optimal symmetric telecloning: the unknown qubit is nonlocally distributed as two best-possible clones $\mathrm{C}_1$ and $\mathrm{C}_2$ while the original qubit at Alice's side is destroyed by the Bell measurement.


The resource state $\ket{\Psi^-_4}_\mathrm{PAC_1C_2}$ provides the nonlocal correlations required to encode the unknown input into Bob's local three qubits (A, C\textsubscript{1} and C\textsubscript{2}) after Alice's BSM. This operational capability is directly rooted in its entanglement structure: from the Schmidt decomposition in Eq. (2), with $\langle \phi_0|\phi_1\rangle_\mathrm{AC_1C_2} = 0$, the reduced state of qubit P is maximally mixed, $\rho_P = \frac{1}{2}\mathbb{I}$, yielding a von Neumann entropy $S(\rho_P)=1$. The state therefore carries exactly one ebit across the P versus (A, C\textsubscript{1}, C\textsubscript{2}) bipartition, representing the minimal entanglement cost required for symmetric $1 \rightarrow 2$ telecloning. It is lower than that required by distributing two copies via independent quantum teleportation.

\section{Experiment}

Our experiment uses two independent integrated photonic chips (Fig. 1b). These devices are fabricated on the silicon-on-insulator (SOI) platform, leveraging its high refractive index contrast and CMOS compatibility to achieve dense component integration with high phase stability, see Supplementary Note 1 for details. Both chips are pumped by a pulsed laser ($\lambda$ = 1550.12 nm) with a repetition rate of 500 MHz, 1.1 ps pulse duration and 9 mW average power. Chip 1 (Alice) contains a nanowire photon-pair source S\textsubscript{1}.  Through the spontaneous four-wave mixing (SFWM) process, S\textsubscript{1} produces a nondegenerate photon pair: a trigger photon, T, at 1554.94 nm and the signal photon, X, at 1545.32 nm to be telecloned. The two photons are separated by wavelength using an on-chip asymmetric Mach-Zehnder interferometer (AMZI\textsubscript{1}), which functions as a wavelength demultiplexer. Hereafter we denote the logical single-photon basis of each qubit \(Q\in\{\mathrm{X,P,A,C_1,C_2}\}\) by \(\ket{0}_Q\) and \(\ket{1}_Q\).  Note that the labels \(\mathrm{X}_0\) and \(\mathrm{X}_1\) in Fig.~1b (and similarly \(\mathrm{P_0,P_1,A_0,A_1,C_{1-0},C_{1-1},C_{2-0},C_{2-1}}\)) refer to physical path (waveguide) indices on the chips rather than abstract computational basis states.  In the experiment the logical states \(\ket{0}_Q\) and \(\ket{1}_Q\) are implemented by photon occupation of the corresponding spatial modes, i.e. presence of the photon in path \(Q_0\) or \(Q_1\), respectively.

On Chip 2 (Bob), we prepare the four-photon entangled state $\ket{\Psi^-_4}_\mathrm{PAC_1C_2}$ (Eq. \eqref{eq:Psi4}) required for the symmetric $1\rightarrow2$ quantum telecloning protocol. The four photons are generated by two nanowire photon-pair sources, S\textsubscript{2} and S\textsubscript{3}, which exhibit high indistinguishability (see Supplementary Note 2 for details). The photon pairs emitted from each source are firstly separated by wavelength using AMZI\textsubscript{2} and AMZI\textsubscript{3}. Following this, modes A and C\textsubscript{1} are swapped. Each of the four resulting modes is then directed through a 2$\times$2 multimode-interference (MMI) coupler that act as a beam-splitter (BS) and subsequently through a routing network. The target resource state $\ket{\Psi^-_4}_\mathrm{PAC_1C_2}$ (Eq. \eqref{eq:Psi4}) is prepared via the detection of four photons. The detailed derivation of the generation process of $\ket{\Psi^-_4}_\mathrm{PAC_1C_2}$ is given in Methods.

Photon P is routed from Chip 2 to Chip 1 and, together with X, is subjected to the on-chip BSM that teleports X. To ensure photons P and X arrive at the BSM module simultaneously and thus interfere, we insert an optical delay line (ODL) into the pump path to Chip 1, which pumps source S\textsubscript{1}. A polarization controller (PC) is used to preserve the polarization state of photon P during its transmission through the fiber. The on-chip BSM module on Chip 1 comprises four tunable BSs (BS\textsubscript{1-4}) and a path-mode parity sorter (PPS). The detailed description of the on-chip BSM is given in Methods. Our on-chip BSM discriminates two Bell states: \(\ket{\Phi^+}\) and \(\ket{\Psi^+}\) on Chip 1. We record the conditional states of C\textsubscript{1} and C\textsubscript{2} corresponding to the BSM outcomes \(\ket{\Phi^+}\) and \(\ket{\Psi^+}\). Their fidelities are then evaluated with respect to the target state $\ket{\varphi}$, after numerically applying the corresponding $\sigma_y$ and $\sigma_z$ unitary corrections to the reconstructed states in data post-processing.

\begin{figure*}
\includegraphics[width=16cm]{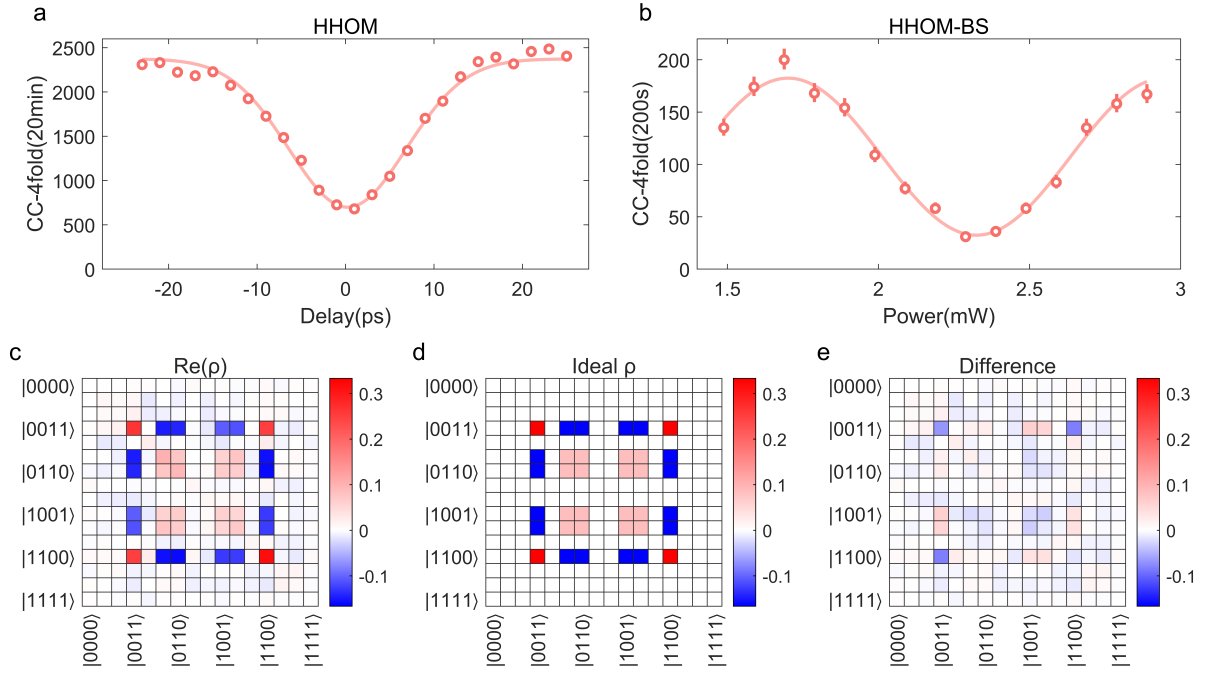}
\caption{The experimental results for the four-photon entangled state $\ket{\Psi^-_4}$ and the BSM. \textbf{a}, Interchip HHOM interference result. After transmitting photon P to Chip 2, we scan the optical delay line to measure two-photon interference between X and P while registering four-fold coincidences (Trigger, X, P, and A). The HHOM visibility is 70.55 $\pm$ 0.36\%. \textbf{b}, Dependence of four-photon coincidence rate on the phase of the PPS in the BSM. The transmission of BS\textsubscript{1}–BS\textsubscript{4} are fixed at 50:50. The interferometer phase is varied by tuning the voltage of the phase shifter PS\textsubscript{5}. The interference visibility is 70.04 $\pm$ 2.12\%. \textbf{c},  
Real part of the reconstructed density matrix of the quantum state, $\ket{\Psi^-_4}$ for the interchip configuration. \textbf{d}, Real part of the ideal target density matrix. \textbf{e}, Difference between the reconstructed and ideal density matrices. The experimentally reconstructed state exhibits a fidelity of $F = 81.12 \pm 0.26\%$ with respect to the ideal $\ket{\Psi^-_4}$.}
\end{figure*}

\section{Results}
We first characterize the performance of the integrated components on the chip (Fig. 1b). Each waveguide source (S\textsubscript{1}-S\textsubscript{3}) delivers an average photon-pair rate of approximately $215 \text{ kHz}$. The edge couplers exhibit a chip-to-fiber coupling efficiency of $\approx 70\%$, which enables the detection of six-photon coincidences at a rate of 10.5 counts per hour for the interchip configuration. For chip-to-chip transmission, on-chip Polarization Beam Rotator Splitters/Combiners (PBRS/PBRC) are utilized to convert between on-chip path modes and fiber polarization modes. The PBRS has the typical insertion and conversion losses of about 0.1~dB and 0.2~dB, respectively, while maintaining a polarization crosstalk better than 18.5~dB. For the PBRC, the typical TE-to-TE insertion loss and TE-to-TM conversion loss are 0.43~dB and 0.25~dB, respectively. Further details for the sources, PBRS/Cs, edge couplers and other integrated components are provided in Supplementary Note 1.

Building upon the characterized performance of the integrated components, we proceed to a series of system-level tests for quantum telecloning.
We sequentially characterize the BSM module via heralded Hong-Ou-Mandel (HHOM) interference and its phase dependence, and assess the purity of the multipartite resource state using complete quantum state tomography.

We measure the HHOM interference between photons X and P. The common pump is split to drive both chips, with photon P from Chip 2 coupled into Chip 1 via an optical fiber. To observe the interference, photons X and P are initialized in state $\ket{0}$, with each photon occupying its respective on-chip path X$_0$ or P$_0$ (Fig. 1b). We configure the interferometric network to route both photons toward the same on-chip beamsplitter (BS$_3$). Specifically, by setting the phases of BS$_1$ to $0$ and BS$_2$ to $\pi$, photon X remains in path X$_0$ (state $\ket{0}_\mathrm{X}$), while photon P is flipped to path P$_1$ (state $\ket{1}_P$). The PPS then acts as a mode-exchanger, routing the photon from P$_1$ into path X$_1$. Consequently, photons X and P are guided to the two input ports of BS$_3$ (with its phase set to $\pi/2$), which serves as the 50:50 interference coupler. By scanning the ODL and recording four-fold coincidences between the trigger, photon $A$, and the detectors at the output ports of X$_0$ and X$_1$, we observe the HHOM dip shown in Fig. 2a. The measured visibility of $70.55 \pm 0.36\%$ enables the realization of interchip BSM.

\begin{figure*}
\includegraphics[width=16cm]{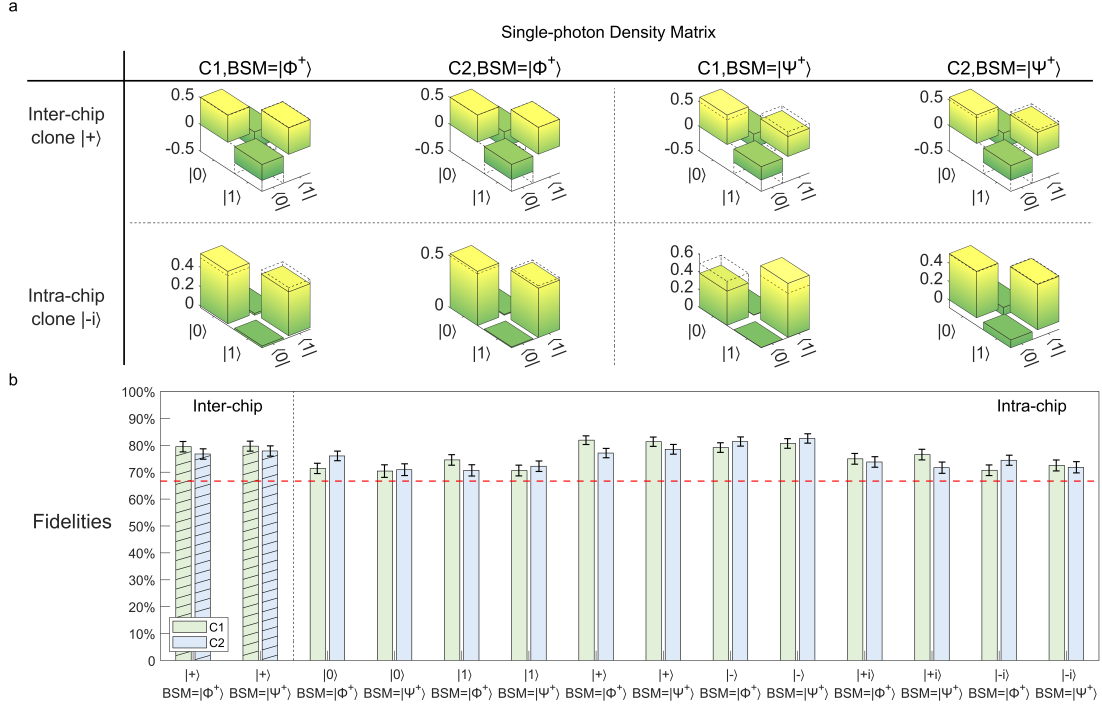}
\caption{Teleclone states and fidelities. \textbf{a}, Experimentally reconstructed single-qubit reduced density matrices for photons C\textsubscript{1} and C\textsubscript{2}, conditioned on the BSM outcome, for two representative input states $\ket{+}_\mathrm{X}=\frac{\ket{0}+\ket{1}}{\sqrt{2}}$ (interchip) and $\ket{-i}_\mathrm{X}=\frac{\ket{0}-i\ket{1}}{\sqrt{2}}$ (intrachip). C\textsubscript{1} and C\textsubscript{2} are the optimal clones for input X. For each input we display the real part of the experimentally reconstructed density matrices for C\textsubscript{1} and C\textsubscript{2}. The corresponding ideal theoretical density matrices are shown with grid bars.
\textbf{b}, Fidelities of the single-photon outputs to the input state $\ket{\varphi}_\mathrm{X}$: bars correspond to C\textsubscript{1}(green), C\textsubscript{2}(blue) for the different input states and conditional BSM outcomes. Error bars are obtained by Monte Carlo sampling of the raw counts assuming Poissonian noise. All measured fidelities of C\textsubscript{1} and C\textsubscript{2} exceed the classical cloning bound 2/3. }
\end{figure*}

Furthermore, we measure the dependence of the HHOM interference on the phase controlled by the phase shifter PS\textsubscript{5} in the BSM. With BS\textsubscript{1}-BS\textsubscript{4} fixed at 50:50, we vary the interferometer phase by changing the voltage of PS\textsubscript{5} and record four-fold coincidences between the trigger, input photon X on port X$_0$, photon P on port P$_0$ and ancillary photon A. The resulting modulation of the four-photon coincidence rate is plotted in Fig. 2b, and a sinusoidal fit yields an interference visibility V = 70.04 $\pm$ 2.12\%, indicating the indistinguishability of photons X and P. Detailed information regarding the phase characterization of the BSM module is provided in Supplementary Note 3.

We perform complete quantum state tomography \cite{PhysRevA.64.052312} to characterize the four-photon resource \(\ket{\Psi^-_4}_\mathrm{PAC_1C_2}\) in two configurations:(i) the intrachip configuration, where all photons are generated and measured on Chip 2, and (ii) the interchip configuration, where photon P is transmitted to Chip 1 from Chip 2. The reconstructed density matrices for the interchip state are presented in Fig. 2c-e. These include the real part of the experimentally obtained (Fig. 2c) and ideal (Fig. 2d) density matrices, with their differences illustrated in Fig. 2e. We obtain a state fidelity of 81.12 $\pm$ 0.26\% for the interchip configuration. The complete results for both the interchip and the intrachip configurations are provided in Supplementary Note 4. High fidelity of the resource state enables clone-state fidelity above the classical bound (66.7\%).

Beyond its overall fidelity, the reconstructed density matrix allows us to further analyze the correlation structure of the four-photon resource state \(\ket{\Psi^-_4}_\mathrm{PAC_1C_2}\). From the reduced density matrix of photon P, we extract a von Neumann entropy $S(\rho_P)=0.9949 \pm 0.0004$, indicating that the local state of photon 
P is very close to maximally mixed. This behavior is consistent with the structure of the ideal telecloning resource, where photon P forms an effective entangled channel with the remaining subsystem $\mathrm{(A,C_1,C_2)}$. In the ideal limit, this bipartition corresponds to one ebit of shared entanglement, which suffices to distribute two optimal clones \cite{PhysRevA.59.156}.

Fig. 3 summarizes the teleclone states and their fidelities. We record sixfold coincidences among the detectors for T, X, P, A, C\textsubscript{1} and C\textsubscript{2} and sort the events into two categories corresponding to BSM outcomes $\ket{\Phi^+}$ and $\ket{\Psi^+}$. For each we perform quantum state tomography on the clone photons and obtain the single-photon reduced density matrices. Fig. 3a shows the real parts of the reconstructed density matrices for C\textsubscript{1} and C\textsubscript{2} conditioned on the BSM outcome, displayed for the two input states $\frac{\ket{0}+\ket{1}}{\sqrt{2}}$ (interchip) and $\ket{\varphi}_\mathrm{X}=\frac{\ket{0}-i\ket{1}}{\sqrt{2}}$ (intrachip). Fig. 3b presents the fidelities of the outputs to the input states $\ket{\varphi}_\mathrm{X}$ — bars correspond to photons C\textsubscript{1}(green), C\textsubscript{2}(blue) for the different inputs and conditional BSM outcomes. The average fidelity of the two clones, obtained by averaging over all measured configurations, is 78.45 $\pm$ 1.39\% for the interchip case and 75.25 $\pm$ 4.16\% for the intrachip case. All fidelities shown in Fig. 3b exceed the classical bound of 66.7\%, indicating the quantum nature of the telecloning process. The corresponding tomography results and fidelity analysis are provided in Supplementary Note 5.


\section{Conclusion}
We have demonstrated the optimal symmetric 1$\rightarrow$2 quantum telecloning across two spatially separated silicon photonic chips. By generating and characterizing the multipartite resource $\ket{\Psi^-_4}_\mathrm{PAC_1C_2}$ on Chip 2, and performing an interchip BSM, we have realized optimal symmetric telecloning. The experimentally reconstructed states of the clone modes show fidelities above the classical cloning bound, confirming the quantum nature of the optimal symmetric telecloning. 

Advanced quantum information processing relies on transferring photonic qubits across different integrated quantum chips. Previous works have achieved distribution of two-photon \cite{zheng2023multichip} and four-photon \cite{d53g-v8q6,Liu2025ChipToChip} entangled states. In this work, we realized the quantum teleconing in a six-photon experiment, marking as a decisive advancement in photonic QIP.

Despite these important results, several aspects warrant further improvement. Specifically, an active feedforward operation has yet to be implemented to rotate the clone output into the target state based on the BSM outcome. In the current study, we rely on quantum state tomography performed on post-selected coincidences correlated with the BSM results. To overcome this limitation, high-speed active feedforward could be implemented by integrating fast electro-optic modulators and superconducting single-photon detectors on a thin-film lithium niobate platform \cite{Lomonte:2021hle}, which may allow for on-chip feed-forward operation. The reliable on-chip generation and full characterization of the four-photon $\ket{\Psi^-_4}_\mathrm{PAC_1C_2}$ resource, together with a six-photon interchip telecloning demonstration, pave practical routes toward integrated quantum-network hardware \cite{5c5v-cw8c}. Beyond enabling scalable telecloning modules for multi-node distributed quantum information processing, these advances point toward concrete applications including multi-party quantum communication \cite{wehner2018quantum}, and distributed quantum sensing \cite{Liu2021Distributed,zhang2021distributed}.

\section{Method}
\subsection{\texorpdfstring{Generate $\ket{\Psi^-_4}$ state for telecloning}{}}

As shown in Fig. 1b, Chip 2 contains two nanowire photon-pair sources, $\mathrm{S}_2$ and $\mathrm{S}_3$, whose outputs are routed through asymmetric Mach–Zehnder interferometers (AMZI\textsubscript{2} and AMZI\textsubscript{3}). The four output modes after the AMZIs are labelled P, A, $\mathrm{C}_1$ and $\mathrm{C}_2$. Retaining only the terms that contribute to four-photon coincidence events yields the following normalized superposition:

\begin{align}
\label{eq:init}
|\Psi_{\mathrm{init}}\rangle_{\mathrm{PAC_1C_2}}
={}&\frac{1}{\sqrt{3}}\Bigg[
e^{i\varphi}
a^\dagger_{\mathrm{P},0}
a^\dagger_{\mathrm{A},0}
a^\dagger_{\mathrm{C}_1,0}
a^\dagger_{\mathrm{C}_2,0}
\nonumber\\
&\quad
+\frac{e^{i2\varphi}}{2}
\left(
a^\dagger_{\mathrm{P},0}
a^\dagger_{\mathrm{A},0}
\right)^2
\nonumber\\
&\quad
+\frac{1}{2}
\left(
a^\dagger_{\mathrm{C}_1,0}
a^\dagger_{\mathrm{C}_2,0}
\right)^2
\Bigg]
|\mathrm{vac}\rangle .
\end{align}

where $a^\dagger_{\mathrm{M},i}$ denotes the creation operator for photon M in mode $i$. The parameter $\varphi$ represents the relative phase of the pump between sources $\mathrm{S}_2$ and $\mathrm{S}_3$, which is controlled by the phase shifter $\mathrm{PS}_0$. In our experiment, we set $\varphi = \pi$.

We then swap the modes of photons A and $\mathrm{C}_1$. Each of the four photons P, A, $\mathrm{C}_1$ and $\mathrm{C}_2$ is subsequently sent through a 50:50 beam splitter, so that the state becomes

\begin{align}
|\Psi_{1}\rangle_{\mathrm{PAC_1C_2}}
={}&\frac{1}{\sqrt{3}}\Bigg[
-\frac{a^\dagger_{\mathrm{P},0}+a^\dagger_{\mathrm{P},1}}{\sqrt{2}}
 \frac{a^\dagger_{\mathrm{C}_1,0}+a^\dagger_{\mathrm{C}_1,1}}{\sqrt{2}}
\nonumber\\
&\qquad\otimes
 \frac{a^\dagger_{\mathrm{A},0}+a^\dagger_{\mathrm{A},1}}{\sqrt{2}}
 \frac{a^\dagger_{\mathrm{C}_2,0}+a^\dagger_{\mathrm{C}_2,1}}{\sqrt{2}}
\nonumber\\
&\quad
+\frac{1}{2}
\left(
 \frac{a^\dagger_{\mathrm{P},0}+a^\dagger_{\mathrm{P},1}}{\sqrt{2}}
 \frac{a^\dagger_{\mathrm{C}_1,0}+a^\dagger_{\mathrm{C}_1,1}}{\sqrt{2}}
\right)^2
\nonumber\\
&\quad
+\frac{1}{2}
\left(
 \frac{a^\dagger_{\mathrm{A},0}+a^\dagger_{\mathrm{A},1}}{\sqrt{2}}
 \frac{a^\dagger_{\mathrm{C}_2,0}+a^\dagger_{\mathrm{C}_2,1}}{\sqrt{2}}
\right)^2
\Bigg]
|\mathrm{vac}\rangle. \nonumber
\end{align}

Next, we relabel the paths of different modes by swapping the physical waveguides, specifically $a^\dagger_\mathrm{P,1}\leftrightarrow a^\dagger_\mathrm{A,0}$ and $a^\dagger_\mathrm{C_1,0}\leftrightarrow a^\dagger_\mathrm{C_2,1}$. Applying these relabellings to the state above yields

\begin{align}
|\Psi_{2}\rangle_{\mathrm{PAC_1C_2}}
={}&\frac{1}{\sqrt{3}}\Bigg[
-\frac{a^\dagger_{\mathrm{P},0}+a^\dagger_{\mathrm{A},0}}{\sqrt{2}}
 \frac{a^\dagger_{\mathrm{C}_2,1}+a^\dagger_{\mathrm{C}_1,1}}{\sqrt{2}}
\nonumber\\
&\qquad\otimes
 \frac{a^\dagger_{\mathrm{P},1}+a^\dagger_{\mathrm{A},1}}{\sqrt{2}}
 \frac{a^\dagger_{\mathrm{C}_2,0}+a^\dagger_{\mathrm{C}_1,0}}{\sqrt{2}}
\nonumber\\
&\quad
+\frac{1}{2}
\left(
 \frac{a^\dagger_{\mathrm{P},0}+a^\dagger_{\mathrm{A},0}}{\sqrt{2}}
 \frac{a^\dagger_{\mathrm{C}_2,1}+a^\dagger_{\mathrm{C}_1,1}}{\sqrt{2}}
\right)^2
\nonumber\\
&\quad
+\frac{1}{2}
\left(
 \frac{a^\dagger_{\mathrm{P},1}+a^\dagger_{\mathrm{A},1}}{\sqrt{2}}
 \frac{a^\dagger_{\mathrm{C}_2,0}+a^\dagger_{\mathrm{C}_1,0}}{\sqrt{2}}
\right)^2
\Bigg]
|\mathrm{vac}\rangle .\nonumber
\end{align}

By selecting events with one detected photon in each mode P, A, $\mathrm{C}_1$ and $\mathrm{C}_2$ — and renormalizing the resulting components, we obtain:
\begin{align}
\nonumber |\Psi _\mathrm{f}\rangle &=\frac{1}{\sqrt{3}}[-\frac{a^\dagger_\mathrm{P,0} a^\dagger_\mathrm{A,1}+a^\dagger_\mathrm{P,1}a^\dagger_\mathrm{A,0}}{\sqrt{2}} \otimes \frac{a^\dagger_\mathrm{C_1,0} a^\dagger_\mathrm{C_2,1}+a^\dagger_\mathrm{C_1,1}a^\dagger_\mathrm{C_2,0}}{\sqrt{2}}\\
&+ a^\dagger_\mathrm{P,0} a^\dagger_\mathrm{A,0} a^\dagger_\mathrm{C_1,1} a^\dagger_\mathrm{C_2,1} + a^\dagger_\mathrm{P,1} a^\dagger_\mathrm{A,1} a^\dagger_\mathrm{C_1,0} a^\dagger_\mathrm{C_2,0}]\ket{\mathrm{vac}}
\end{align}
which is the resource state $\ket{\Psi^{-}_4}_\mathrm{PAC_1C_2}$ (Eq. \eqref{eq:Psi4}) required for the optimal symmetric telecloning protocol.

\subsection{BSM on Chip 1}

The BSM module on Chip 1 comprises four tunable BSs (BS\textsubscript{1}-BS\textsubscript{4}) and a central PPS (see Fig. 1b, Chip 1). Each tunable BS is implemented as two MMIs with a thermo-optic phase shifter between them; the phase shifters are biased to yield an effective 50:50 splitting (BS\textsubscript{1} and BS\textsubscript{3} set to \(\pi/2\), BS\textsubscript{2} and BS\textsubscript{4} set to \(3\pi/2\)). The central PPS implements the following mapping on the computational basis:
\begin{align}
\nonumber &\ket{00}_\mathrm{XP}=a^\dagger_\mathrm{X,0}a^\dagger_\mathrm{P,0}\ket{\mathrm{vac}}\rightarrow a^\dagger_\mathrm{X,0}a^\dagger_\mathrm{P,0}\ket{\mathrm{vac}}, \\
\nonumber &\ket{01}_\mathrm{XP}=a^\dagger_\mathrm{X,0}a^\dagger_\mathrm{P,1}\ket{\mathrm{vac}}\rightarrow a^\dagger_\mathrm{X,0}a^\dagger_\mathrm{X,1}\ket{\mathrm{vac}}, \\
\nonumber &\ket{10}_\mathrm{XP}=a^\dagger_\mathrm{X,1}a^\dagger_\mathrm{P,0}\ket{\mathrm{vac}}\rightarrow a^\dagger_\mathrm{P,0}a^\dagger_\mathrm{P,1}\ket{\mathrm{vac}}, \\
\nonumber &\ket{11}_\mathrm{XP}=a^\dagger_\mathrm{X,1}a^\dagger_\mathrm{P,1}\ket{\mathrm{vac}}\rightarrow a^\dagger_\mathrm{X,1}a^\dagger_\mathrm{P,1}\ket{\mathrm{vac}}.
\end{align}

Under these operations the Bell inputs \(\ket{\Phi^+}_\mathrm{XP}\) and \(\ket{\Psi^+}_\mathrm{XP}\) transform as follows:

\begin{align}
\ket{\Phi^+}_{\mathrm{XP}}
&\xrightarrow{\mathrm{BS}_1}
\ket{\Phi^+}_{\mathrm{XP}}
\xrightarrow{\mathrm{PPS}}
\ket{\Phi^+}_{\mathrm{XP}}
\nonumber\\
&\xrightarrow{\mathrm{BS}_3}
\frac{1}{\sqrt{2}}
\left(
a^\dagger_{\mathrm{X},0}a^\dagger_{\mathrm{P},0}
+
a^\dagger_{\mathrm{X},1}a^\dagger_{\mathrm{P},1}
\right)
\ket{\mathrm{vac}},
\\[3pt]
\ket{\Psi^+}_{\mathrm{XP}}
&\xrightarrow{\mathrm{BS}_2}
\ket{\Phi^-}_{\mathrm{XP}}
\xrightarrow{\mathrm{PPS}}
\ket{\Phi^-}_{\mathrm{XP}}
\nonumber\\
&\xrightarrow{\mathrm{BS}_4}
\frac{1}{\sqrt{2}}
\left(
a^\dagger_{\mathrm{X},0}a^\dagger_{\mathrm{P},1}
+
a^\dagger_{\mathrm{X},1}a^\dagger_{\mathrm{P},0}
\right)
\ket{\mathrm{vac}} .
\end{align}

These transformations yield distinct coincidence signatures between modes X and P: detection events $\ket{00}_\mathrm{XP}$ or $\ket{11}_\mathrm{XP}$ identify the \(\ket{\Phi^+}_\mathrm{XP}\) outcome, whereas detection of $\ket{01}_\mathrm{XP}$ or $\ket{10}_\mathrm{XP}$ identify the \(\ket{\Psi^+}_\mathrm{XP}\) outcome.

The remaining Bell states \(\ket{\Phi^-}_\mathrm{XP}\) and \(\ket{\Psi^-}_\mathrm{XP}\) evolve according to:
\begin{align}
&\ket{\Phi^-}_\mathrm{XP} \xrightarrow{\mathrm{BS_1}} \ket{\Psi^+}_\mathrm{XP} \xrightarrow{\mathrm{PPS}} \frac{1}{\sqrt{2}}(a^\dagger_\mathrm{X,0}a^\dagger_\mathrm{X,1}+a^\dagger_\mathrm{P,0}a^\dagger_\mathrm{P,1})\ket{\mathrm{vac}}\\
&\ket{\Psi^-}_\mathrm{XP} \xrightarrow{\mathrm{BS_2}} \ket{\Psi^-}_\mathrm{XP} \xrightarrow{\mathrm{PPS}} \frac{1}{\sqrt{2}}(a^\dagger_\mathrm{X,0}a^\dagger_\mathrm{X,1}-a^\dagger_\mathrm{P,0}a^\dagger_\mathrm{P,1})\ket{\mathrm{vac}}
\end{align}

In both cases the two photons emerge in the same output port of the PPS (modes XX or PP) and therefore do not contribute to two-mode coincidences between X and P. Consequently, \(\ket{\Phi^-}_\mathrm{XP}\) and \(\ket{\Psi^-}_\mathrm{XP}\) are not distinguished by our coincidence-based heralding scheme.

\appendix
\section{Supplementary Note}
\subsection{Supplementary Note 1 -- Design and Characterization of the Integrated Photonic Circuit}

The integrated photonic circuit (PIC) utilized in this study is fabricated via the Advanced Micro Foundry (AMF) Multi-Project Wafer (MPW) service, employing a high-performance 220 nm Silicon-on-Insulator (SOI) platform. The device layer comprises a 220 nm thick monocrystalline silicon film atop a 3 $\mu$m thick buried oxide (BOX) layer, providing robust optical confinement and minimizing leakage into the silicon substrate. 

The fundamental building blocks of the circuit are strip waveguides with a nominal width of 500 nm. This geometry is optimized to satisfy the single-mode condition for the transverse electric (TE) mode while achieving precise dispersion engineering within the C-band. These waveguides function both as low-loss interconnects and as the nonlinear medium for Spontaneous Four-Wave Mixing (SFWM). To facilitate the six-photon experiment, a high-brightness photon pair source is required. The sources exhibit an average coincidence rate of about 215 kHz. The four-fold coincidence counting rate for the $\ket{\Psi^-_4}_\mathrm{P, A, C_1, C_2}$ state is about 36.4~Hz for the interchip case. The four-photon coincidence rate between the trigger and photons X, P, and A in the Bell-state measurement is 6.14~Hz, while the final inter-chip six-photon coincidence rate is around 10 per hour. A more detailed analysis of the loss and counts is provided in Supplementary Note 6.

\begin{figure}[H]
    \centering
    \includegraphics[width=0.5\textwidth]{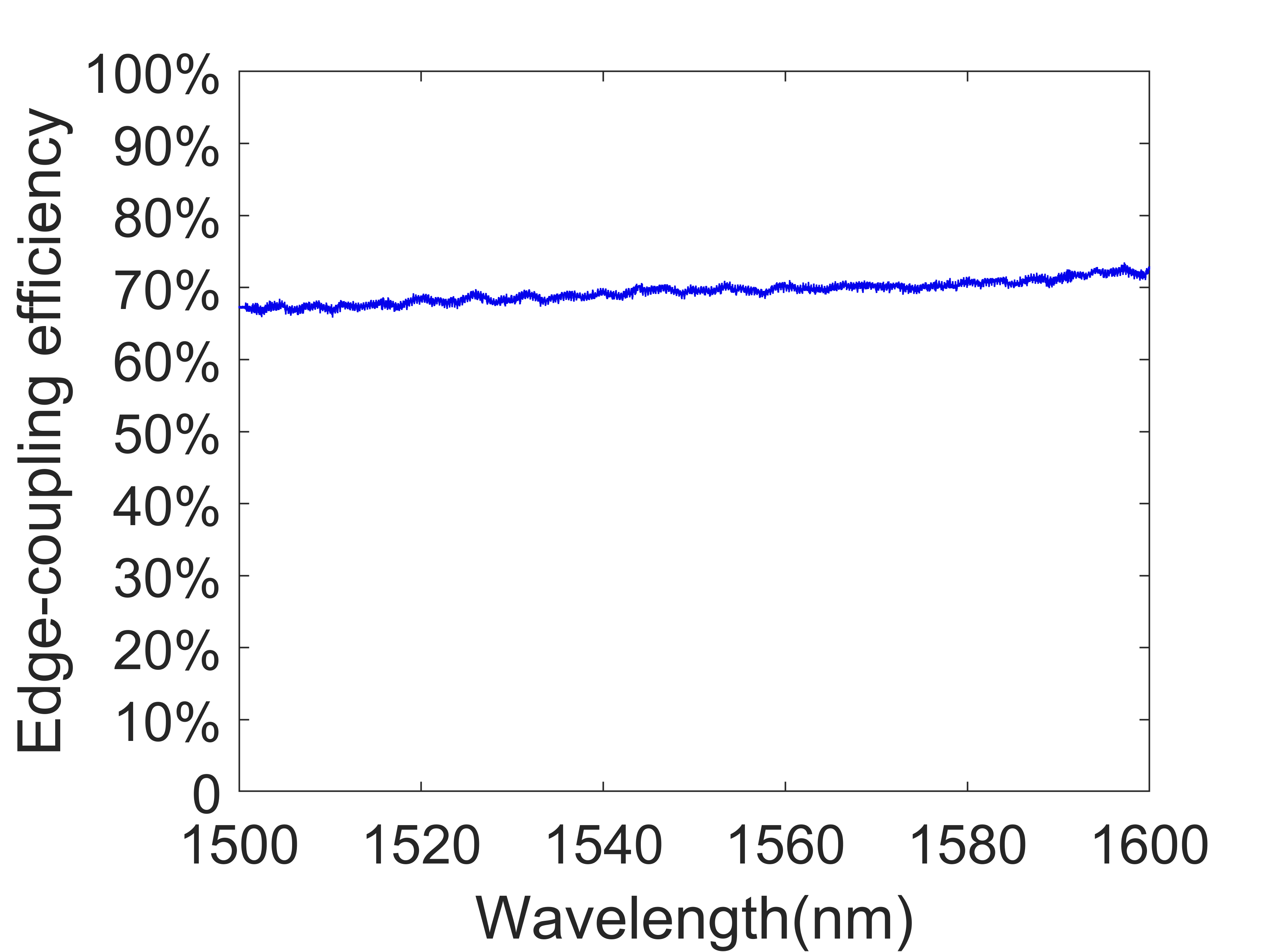}
    \caption{Measured chip-to-fiber coupling efficiency of the edge coupler from 1500 nm to 1600 nm. An average efficiency of approximately 70\% is achieved across the range with the application of refractive index matching fluid.}
    \label{edge_coupler}
\end{figure}

Optical interfacing between the PIC and external fiber arrays is implemented through a hybrid coupling strategy. Grating couplers, optimized for 1550 nm, are utilized for near-vertical injection of the pump laser. For the collection of down-converted photons, which require higher efficiency, 
we employ edge couplers based on an inverse taper design. These structures adiabatically expand the mode field to a $1/e^2$ mode field diameter (MFD) of approximately $9\ \mu\text{m} \times 5\ \mu\text{m}$ at the chip facet, enabling efficient and direct coupling with standard single-mode fibers (SMF). To reduce photon loss at the interface between the chip and the fiber array (FA), the refractive index matching liquid ($n = 1.4300 \pm 0.0002$ at room temperature) is applied to the gap. This approach mitigates mode mismatch, allowing the average edge-coupling efficiency per facet to reach approximately 70\%. Fig.~\ref{edge_coupler} shows the measured spectral response of the edge coupler.

For inter-chip transmission, we integrate on-chip polarization beam rotator-combiners (PBRC) and splitters (PBRS) to enable coherent conversion between path and polarization degrees of freedom. The performance specifications of the PBRS and PBRC used in this work are based on the AMF Process Design Kit (PDK) Manual. The PBRS exhibits a TE-to-TE insertion loss of 0.1 to 0.6 dB and a TM-to-TE conversion loss of 0.2 to 1.0 dB. Its polarization isolation capabilities are further characterized by TE and TM polarization crosstalk values of $19.4 \pm 2.9$ dB and $18.5 \pm 2.3$ dB, respectively. For the PBRC, the TE-to-TE insertion loss is specified as $0.43 \pm 0.21$ dB, while the TE-to-TM conversion loss is $0.25 \pm 0.35$ dB.

On-chip signal routing and splitting are performed by Multimode Interference (MMI) couplers. The typical excess loss is about 0.15~dB and the imbalance is about 0.07~dB \cite{PhysRevLett.130.223601}. Reconfigurable phase control is achieved via integrated thermo-optic phase shifters. These consist of metallic titanium nitride (TiN) heaters above the silicon waveguides. To mitigate thermal crosstalk — a critical factor in high-density quantum photonic integration — deep-trench isolation is implemented. 
Trenches with a width of 11 $\mu$m are etched through the SiO$_2$ cladding, the silicon device layer, and the BOX layer into the silicon substrate, and are positioned 4.5 $\mu$m away from the waveguide core. This design reduces the power consumption per phase shifter. It also restricts heat diffusion, ensuring the phase stability of adjacent interferometric arms, which is essential for high-fidelity quantum state preparation.

\subsection{Supplementary Note 2 -- The indistinguishability of photon-pair sources}
The indistinguishability of the two photon sources is essential for generating high-fidelity four-qubit entangled states. We characterize the indistinguishability of sources S\textsubscript{2} and S\textsubscript{3} by measuring two-fold correlation interference. In our experiment (Fig.~\ref{Chip2}), each photon of a generated pair passes through one of two unitary matrices (e.g., U1 or U2 for one photon, and U3 or U4 for the other), resulting in four categories of correlation interference (U1-U3, U1-U4, U2-U3, U2-U4). Each category comprises four distinct patterns, yielding a total of 16 interference curves obtained by varying the interferometer phase via PS\textsubscript{0}. The measured modulations of the two-photon coincidence rates and their corresponding visibilities are shown in Fig.~\ref{Two-photon}. All two-photon visibilities exceed 90\%, confirming that the two sources are highly indistinguishable.

\subsection{Supplementary Note 3 -- The Bell-state measurement module}
To characterize the phases controlled by PS$_5$ and PS$_6$ within the BSM module, we perform four-fold interference measurements. The beam splitters BS$_1$ and BS$_3$ are set to $\pi/2$, while BS$_2$ and BS$4$ are set to $3\pi/2$. This configuration yields a 50:50 splitting ratio. As illustrated in Fig.~\ref{BSM-PS-scheme}, a $|00\rangle_\mathrm{XP}$ state is prepared (with the remaining two photons serving as triggers), which then undergoes the following evolution:

\begin{equation}
\begin{aligned}
|00\rangle
&\xrightarrow{\mathrm{BS}_{1,2}}
\frac{1}{2} (|0\rangle+|1\rangle)(|0\rangle+|1\rangle) \\
&\xrightarrow{\mathrm{PPS,\ post\text{-}select}}
\frac{1}{\sqrt{2}} (|00\rangle+|11\rangle).
\end{aligned}
\end{equation}

Defining the phases of PS$_5$ and PS$6$ as $\phi_1$ and $\phi_2$, the state evolves into:
\begin{equation}
\begin{aligned}
\frac{1}{\sqrt{2}}
\left(
e^{i\Delta\phi}|00\rangle+|11\rangle
\right)
&\xrightarrow{\mathrm{BS}_{3,4}}
\frac{1}{2\sqrt{2}}
\Big[
(e^{i\Delta\phi}+1)(|00\rangle+|11\rangle)
\\
&\qquad+
(e^{i\Delta\phi}-1)(|01\rangle+|10\rangle)
\Big],
\end{aligned}
\end{equation}
where $\Delta\phi=\phi_1-\phi_2$. Consequently, different four-fold coincidence patterns exhibit sinusoidal modulation with respect to the phases controlled by PS$_5$ and PS$_6$. The curve presented in Fig. 2b of the main text shows the modulation of the $|00\rangle$ output as a function of the power applied to PS$_5$. The complete set of measured modulation curves and their corresponding visibilities are provided in Fig.~\ref{BSM-PS}. The average visibilities obtained by varying PS$_5$ and PS$_6$ are $70.59\% \pm 2.31\%$ and $71.89\% \pm 3.55\%$, respectively.

\subsection{Supplementary Note 4 -- Density matrix reconstruction and fidelity analysis of the four-qubit entangled state}
The main text presents the real parts of the reconstructed density matrices for the four-qubit entangled state in the interchip configuration (Fig. 2). Here, we provide the 
complete results of density matrices (real and imaginary) for the intrachip and interchip configurations (Fig. \ref{densityM}). Note that the imaginary components of the ideal density matrix are zero. We further reconstruct the density matrix for the intrachip configuration (Fig. \ref{densityM}a, b). Due to the increased optical loss inherent in the interchip state, the average pump power is adjusted to 9 mW to maintain signal quality, compared to 7 mW used for the intrachip state. This difference in pump power, along with the increased loss, result in a slight variation in state fidelity: we obtain a fidelity of $83.81\% \pm 0.10\%$ for the intrachip state and $81.12\% \pm 0.26\%$ for the interchip state. The reconstructed density matrices, their corresponding ideal cases, and the absolute differences in the real parts for both configurations are presented in Fig. \ref{densityM}.

\subsection{Supplementary Note 5 -- Single-photon state tomography}
We perform state tomography of the teleclone photons C\textsubscript{1}, C\textsubscript{2} and ancilla photon A simultaneously. Specifically, each combination of phase shifter and Mach-Zehnder interferometer (PS$_2$+MZI$_2$, PS$_3$+MZI$_3$, and PS$_4$+MZI$_4$) is simultaneously configured to project onto the measurement basis (X, Y, or Z). Following the BSM, the data are categorized by BSM outcomes and then partitioned according to the specific photon targeted for tomography. 

For the intrachip configuration, we perform state telecloning for six input states—$\ket{0}$, $\ket{1}$, $\ket{+}$, $\ket{-}$, $\ket{+i}$, and $\ket{-i}$ — accumulating total counts of 1506, 1500, 1538, 1150, 1452, and 1053, respectively. The integration times for these states ranged from approximately 36 to 97.5 hours, as detailed in Tables~\ref{tab:cloning_counts1}-\ref{tab:cloning_counts_minus_i}. The variation in integration times mainly arises from drifts in the fiber-to-chip coupling efficiency during prolonged operation. Since multi-fold coincidence counts are more sensitive to the coupling efficiency, fluctuations in the count rates are effectively amplified over long integration periods. The corresponding density matrices are displayed in Figs.~\ref{intra0}-\ref{intrami}. For the interchip configuration, the $\ket{+}$ state is cloned with 1702 counts in 162.5~h (Table \ref{tab:cloning_counts_interchip_plus}), and its density matrix is shown in Fig. \ref{interp}. Since no physical feed-forward rotations are applied to the cloned photons C\textsubscript{1}, and C\textsubscript{2}, we instead perform an inverse rotation on the ideal target density matrix based on the BSM results. Final fidelities are then determined by comparing the experimental data with these rotated ideal states using the maximum likelihood estimation (MLE) technique.

\subsection{Supplementary Note 6 -- Analysis of photon pair generation, loss, and multi-photon coincidence rates}

\subsubsection{Source Brightness and Loss Estimation}

To characterize the performance of the photon sources, we model the photon-pair sources (S\textsubscript{1}–S\textsubscript{3}) in the low-gain regime and retain terms up to second order in the photon-pair generation probability per pulse $p$ for each source. The third-order contributions of each source are neglected, as they do not contribute to the six-photon events in our experimental configuration and are significantly weaker than the second-order emissions in four-fold coincidence counts.

The experimental single-photon counts ($SC$) and coincidence counts ($CC$) for a signal ($s$) and idler ($i$) photon pair are:
\begin{equation} SC_s = DC_s + f[p\eta_s + p^2(2\eta_s-\eta_s^2)]\end{equation}
\begin{equation} SC_i = DC_i + f[p\eta_i + p^2(2\eta_i-\eta_i^2)]\end{equation}
\begin{equation} CC = ACC + pf\eta_s\eta_i\end{equation}
where $f$ is the laser repetition rate, and $\eta_{s,i}$ represents the transmission efficiency (lumping all losses) of the signal and idler paths, respectively. $DC$ and $ACC$ denote the dark counts and accidental coincidences. Given a measured dark count rate of $DC \approx 200\text{ Hz}$ and experimental values for source S\textsubscript{1} ($SC_s = 5.02 \times 10^5\text{ Hz}$, $SC_i = 5.26 \times 10^5\text{ Hz}$, and $CC = 6.1 \times 10^4\text{ Hz}$), we solve these coupled equations to yield a generation probability $p_1 = 0.0087$ and path losses of $\eta_s = -9.36\text{ dB}$ and $\eta_i = -9.15\text{ dB}$

\subsubsection{Multi-port Splitting and System Loss Analysis}
\begin{figure}[H]
    \centering
    \includegraphics[width=0.9\textwidth]{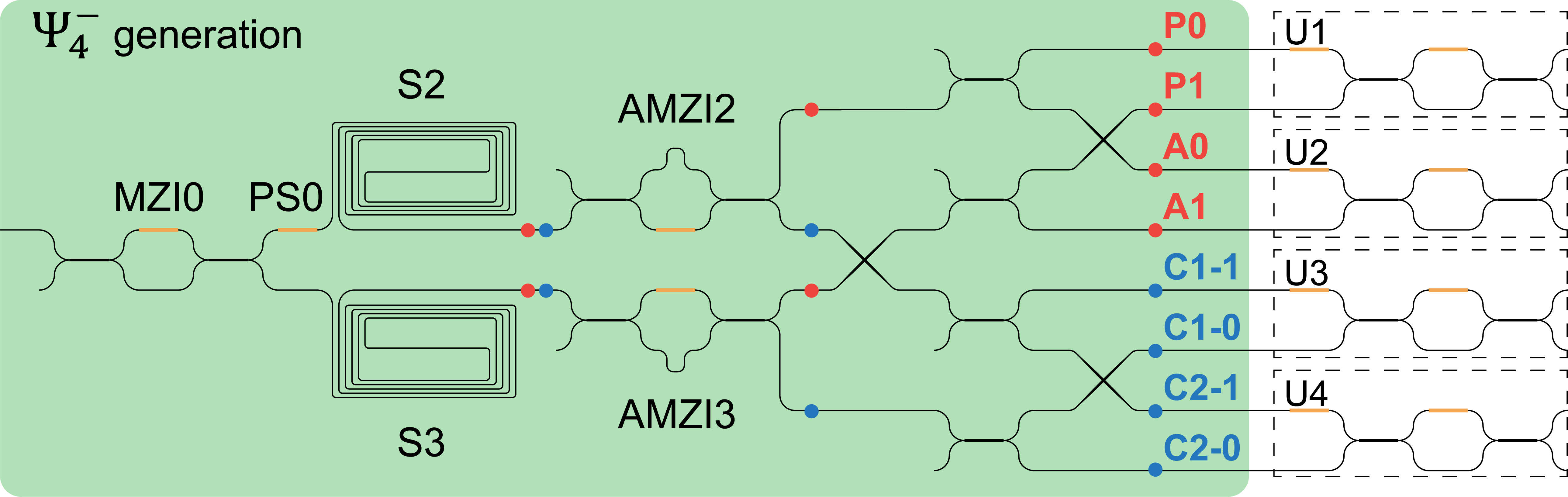}
    \caption{Layout of Chip 2. Unitary modules U\textsubscript{1}-U\textsubscript{4} are used to perform single-qubit state tomography of photons arriving from the photon sources. When source S\textsubscript{2} (resp. S\textsubscript{3}) generates a photon pair, routing directs one photon to either U\textsubscript{1} or U\textsubscript{2} and the other to either U\textsubscript{3} or U\textsubscript{4}. As a result, emission events from S\textsubscript{2} and S\textsubscript{3} are placed into a path superposition. Correlated measurement outcomes across U\textsubscript{1}-U\textsubscript{4} therefore allow us to probe the indistinguishability of S\textsubscript{2} and S\textsubscript{3}.}
    \label{Chip2}
\end{figure}

\begin{figure}[H]
    \centering
    \includegraphics[width=1.0\textwidth]{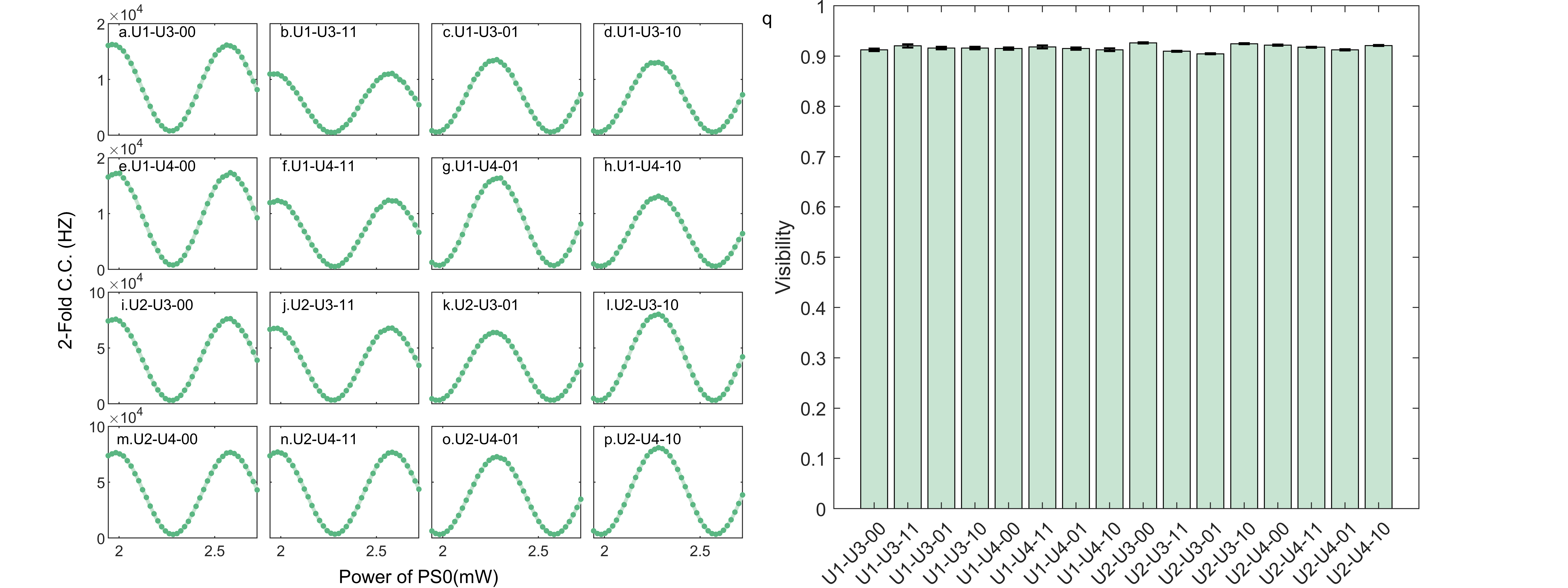}
    \caption{Two-photon correlation interference and visibility characterization. \textbf{a-p}, Measured interference curves for different combinations of unitary matrices: \textbf{a-d}, $\mathrm{U}_1$-$\mathrm{U}_3$, \textbf{e-h}, $\mathrm{U}_1$-$\mathrm{U}_4$, \textbf{i-l}, $\mathrm{U}_2$-$\mathrm{U}_3$, and \textbf{m-p}, $\mathrm{U}_2$-$\mathrm{U}_4$. For each combination, the four panels correspond to the coincidence patterns of $|00\rangle$, $|11\rangle$, $|01\rangle$, and $|10\rangle$, respectively. The modulations were obtained by varying the interferometer phase via PS$_0$. \textbf{q}, Summarized visibilities for all 16 interference patterns, with all values exceeding 90\%.}
    \label{Two-photon}
\end{figure}

\begin{figure}[H]
    \centering
    \includegraphics[width=0.8\textwidth]{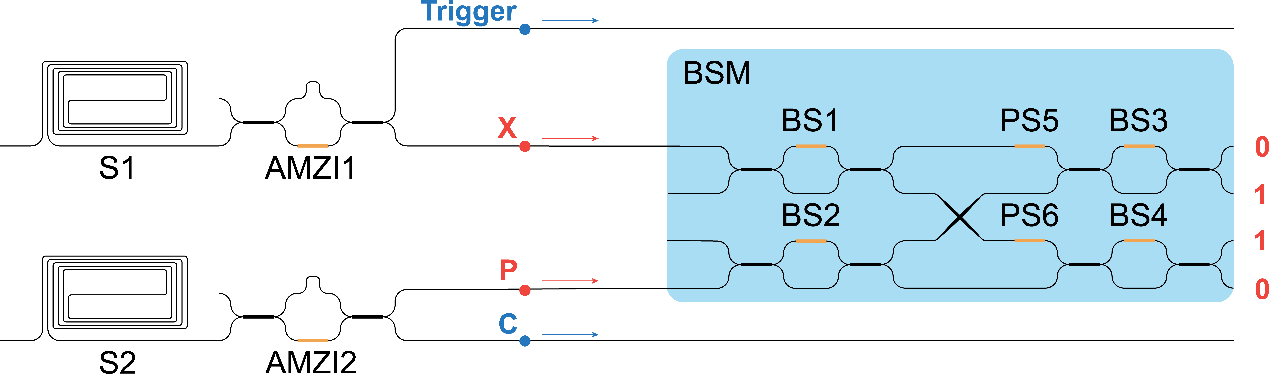}
    \caption{Schematic of the BSM module and phase characterization setup. Photons X and P are initially injected into a 50:50 splitting stage (BS$_1$ and BS$_2$). The resulting state passes through a path-mode parity sorter (PPS). Subsequently, phase shifters PS$_5$ and PS$_6$ introduce relative phases $\phi_1$ and $\phi_2$, followed by a second 50:50 splitting stage (BS$_3$ and BS$_4$) to perform the interference required for phase characterization. This setup allows for the calibration of BSM phases by detecting the sinusoidal modulation of four-fold coincidence counts.}
    \label{BSM-PS-scheme}
\end{figure}

\begin{figure}[H]
    \centering
    \includegraphics[width=0.8\textwidth]{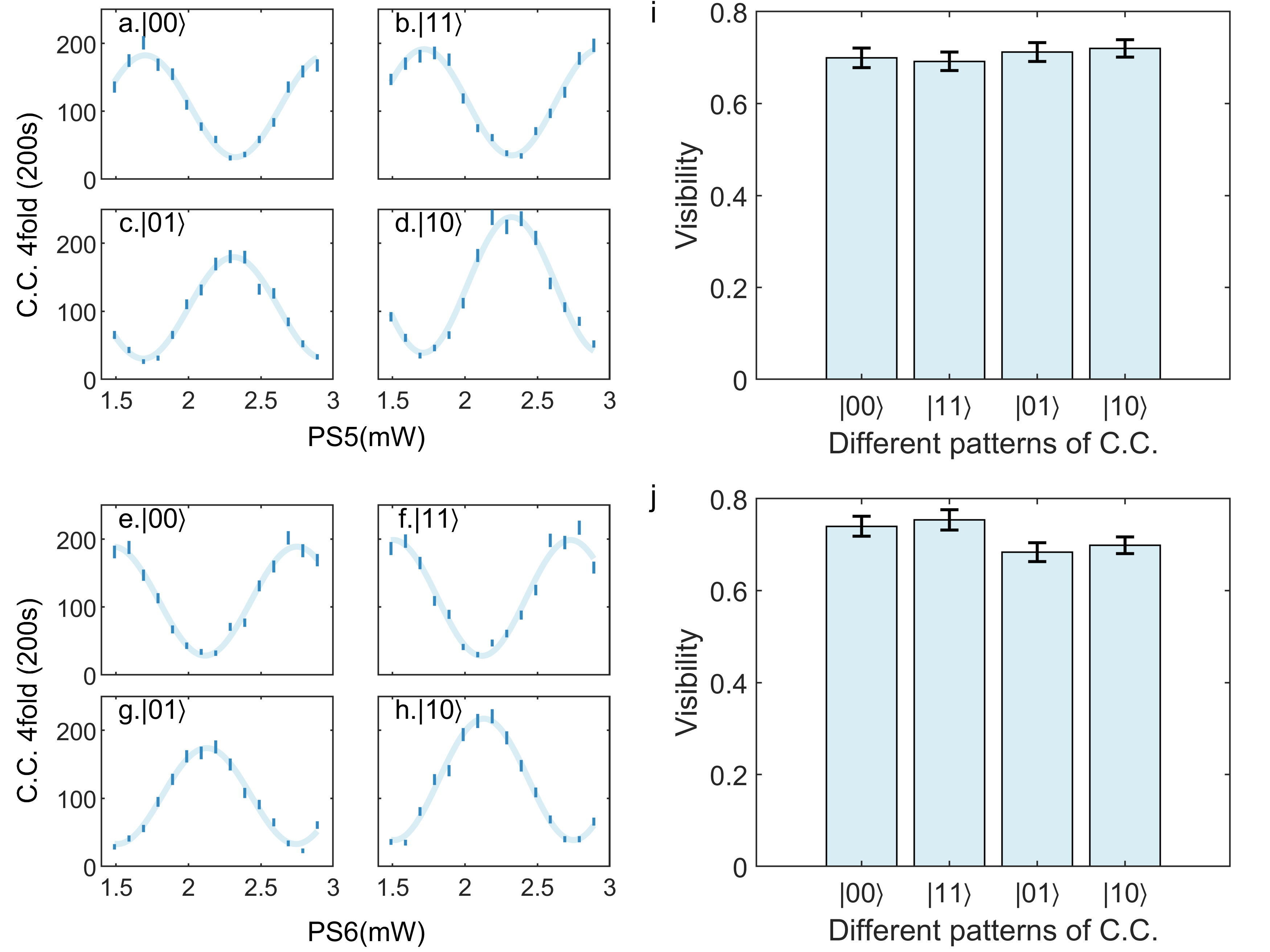}
    \caption{Interference of four-fold coincidence counts and visibility characterization in the BSM module. \textbf{a-h}, Measured four-fold coincidence modulations as a function of BSM phases. \textbf{a-d}, Interference curves obtained by varying the phase $\phi_1$ via PS$_5$, with each panel representing a distinct coincidence pattern ($|00\rangle$, $|11\rangle$, $|01\rangle$, and $|10\rangle$, respectively). \textbf{e-h}, Similar modulations obtained by varying $\phi_2$ via PS$_6$. \textbf{i, j}, Summarized visibilities for the interference patterns controlled by \textbf{i}, PS$_5$ and \textbf{j}, PS$_6$. The high visibilities across all patterns confirm the effective phase control and high indistinguishability within the BSM module.}
    \label{BSM-PS}
\end{figure}

\begin{figure}[H]
    \centering
    \includegraphics[width=1.0\textwidth]{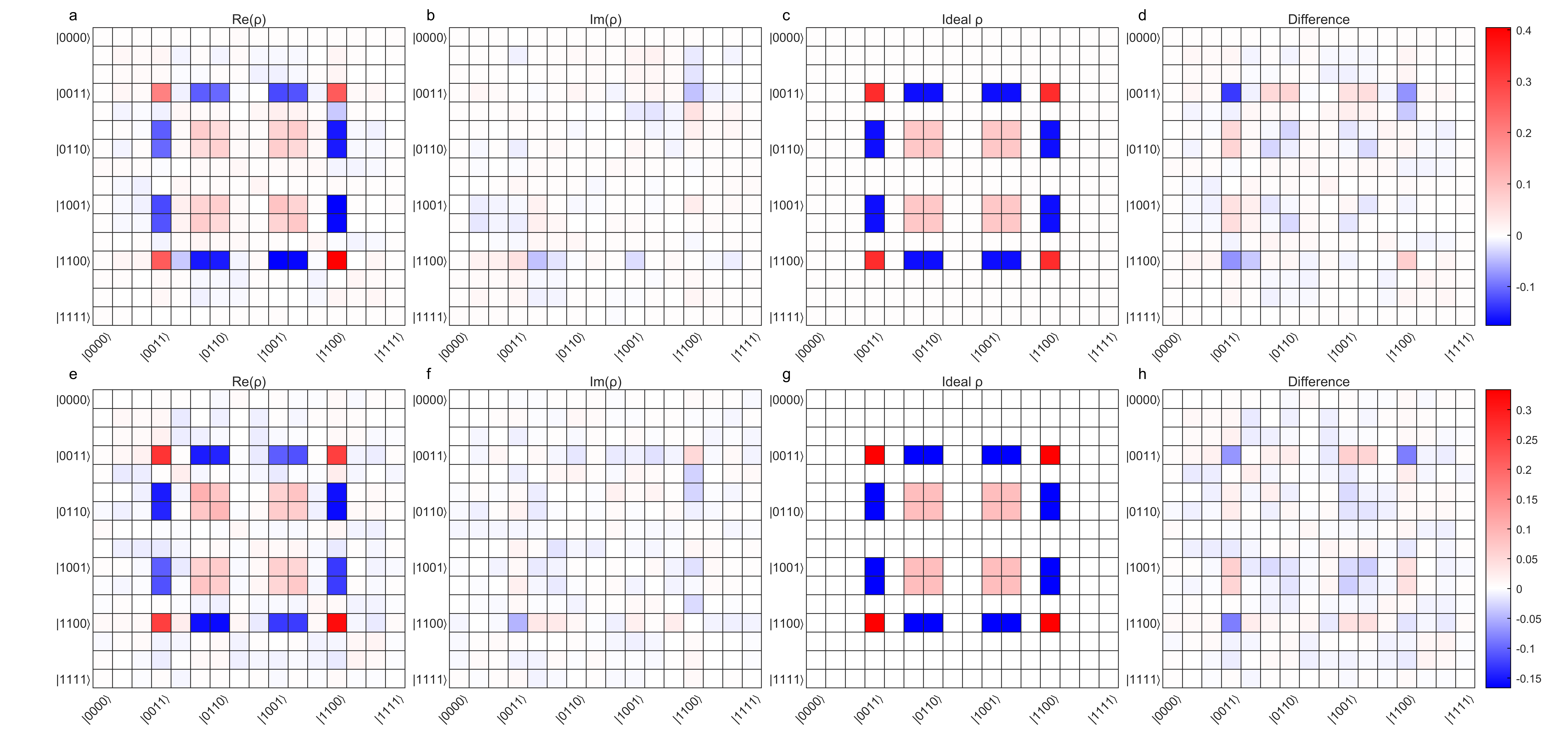}
    \caption{Density matrix reconstruction of the four-qubit entangled state. \textbf{a-d}, Results for the intrachip configuration and \textbf{e-h}, results for the interchip configuration. For each configuration, the panels display the (\textbf{a}, \textbf{e}) real part and (\textbf{b}, \textbf{f}) imaginary part of the reconstructed density matrix, (\textbf{c}, \textbf{g}) the ideal density matrix, and (\textbf{d}, \textbf{h}) the absolute difference between the reconstructed real part and the ideal case. The imaginary parts of the reconstructed matrices are near-zero, consistent with the ideal case.}
    \label{densityM}
\end{figure}

\begin{figure}[H]
    \centering
    \includegraphics[width=0.8\textwidth]{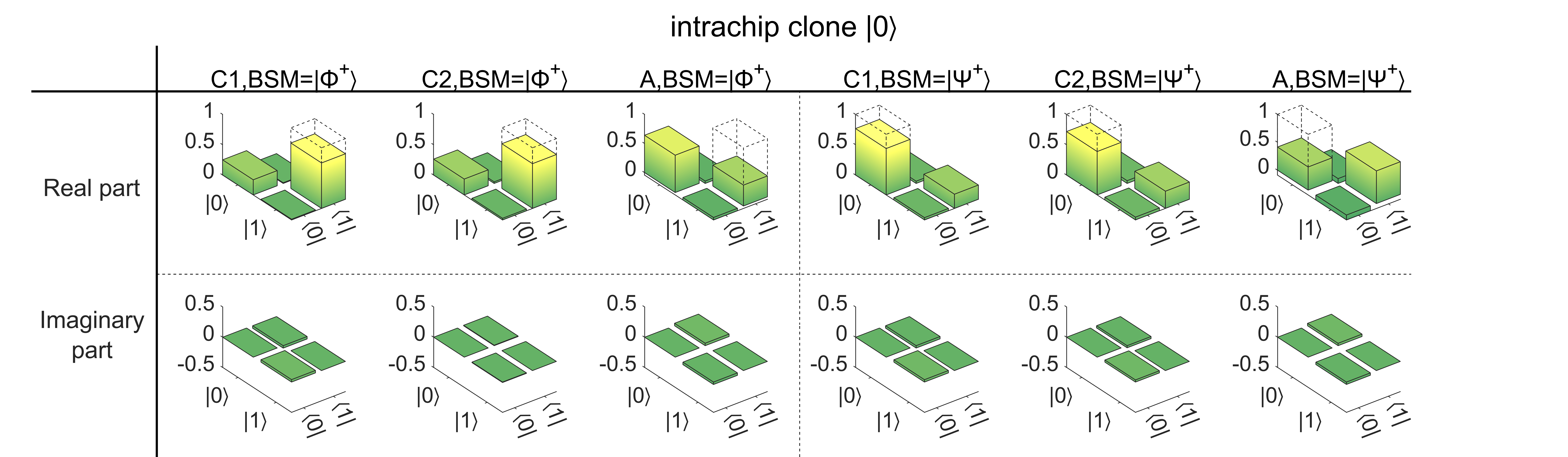}
    \caption{Reconstructed density matrices of photons $\mathrm{C}_1$, $\mathrm{C}_2$, and A for the cloned $|0\rangle$ state (intrachip configuration). The panels display the real and imaginary parts of the density matrices, reconstructed via MLE from parallel single-photon tomography data. The results are categorized by the BSM outcomes: $\ket{\Phi^+}$ and $\ket{\Psi^+}$. Within each category, the reconstructed states for the three individual photons are presented to verify the performance of the optimal symmetric telecloning process.}
    \label{intra0}
\end{figure}

\begin{figure}[H]
    \centering
    \includegraphics[width=0.8\textwidth]{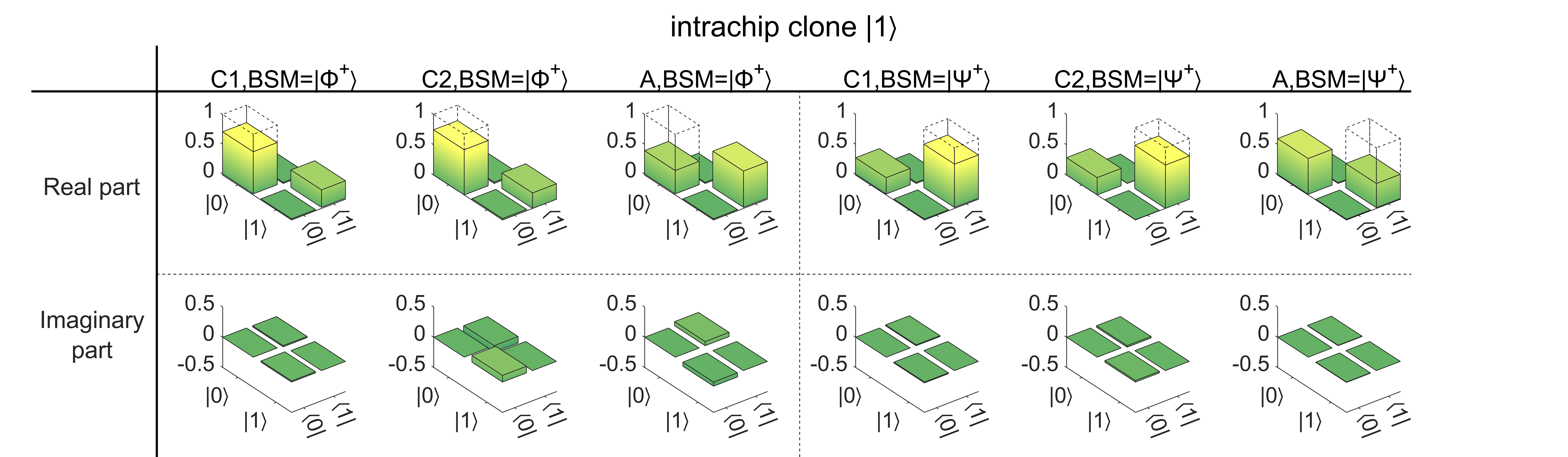}
    \caption{Reconstructed density matrices of photons $\mathrm{C}_1$, $\mathrm{C}_2$, and A for the cloned $|1\rangle$ state (intrachip configuration). The panels display the real and imaginary parts of the density matrices, reconstructed via MLE from parallel single-photon tomography data. The results are categorized by the BSM outcomes: $\ket{\Phi^+}$ and $\ket{\Psi^+}$. Within each category, the reconstructed states for the three individual photons are presented to verify the performance of the optimal symmetric telecloning process.}
    \label{intra1}
\end{figure}

\begin{figure}[H]
    \centering
    \includegraphics[width=0.8\textwidth]{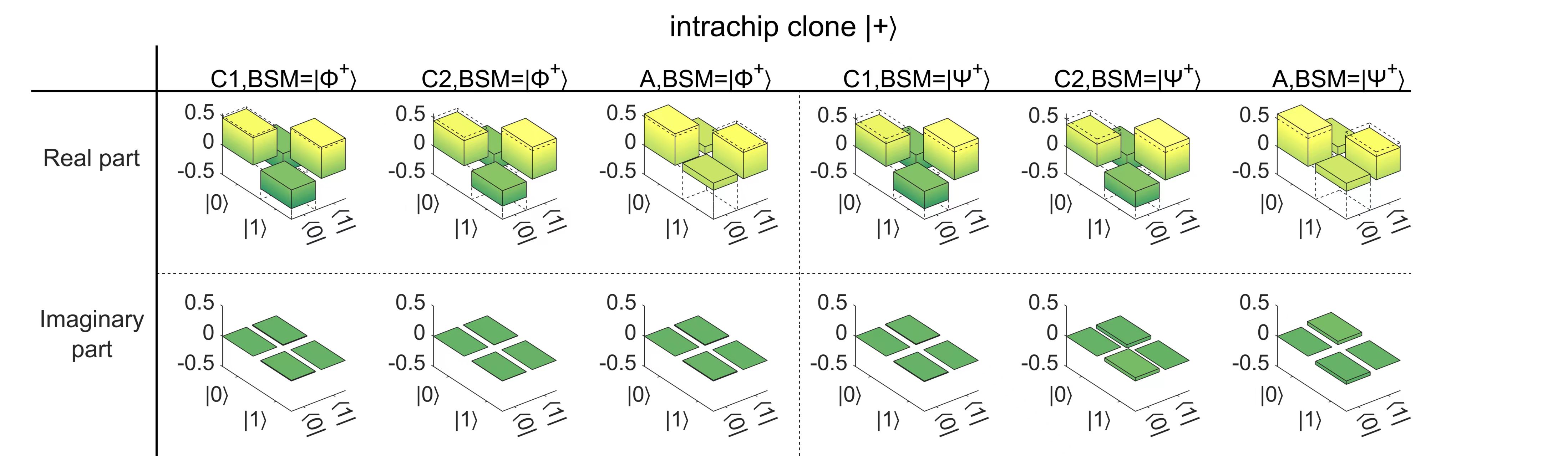}
    \caption{Reconstructed density matrices of photons $\mathrm{C}_1$, $\mathrm{C}_2$, and A for the cloned $|+\rangle$ state (intrachip configuration). The panels display the real and imaginary parts of the density matrices, reconstructed via MLE from parallel single-photon tomography data. The results are categorized by the BSM outcomes: $\ket{\Phi^+}$ and $\ket{\Psi^+}$. Within each category, the reconstructed states for the three individual photons are presented to verify the performance of the optimal symmetric telecloning process.}
    \label{intrap}
\end{figure}

\begin{figure}[H]
    \centering
    \includegraphics[width=0.8\textwidth]{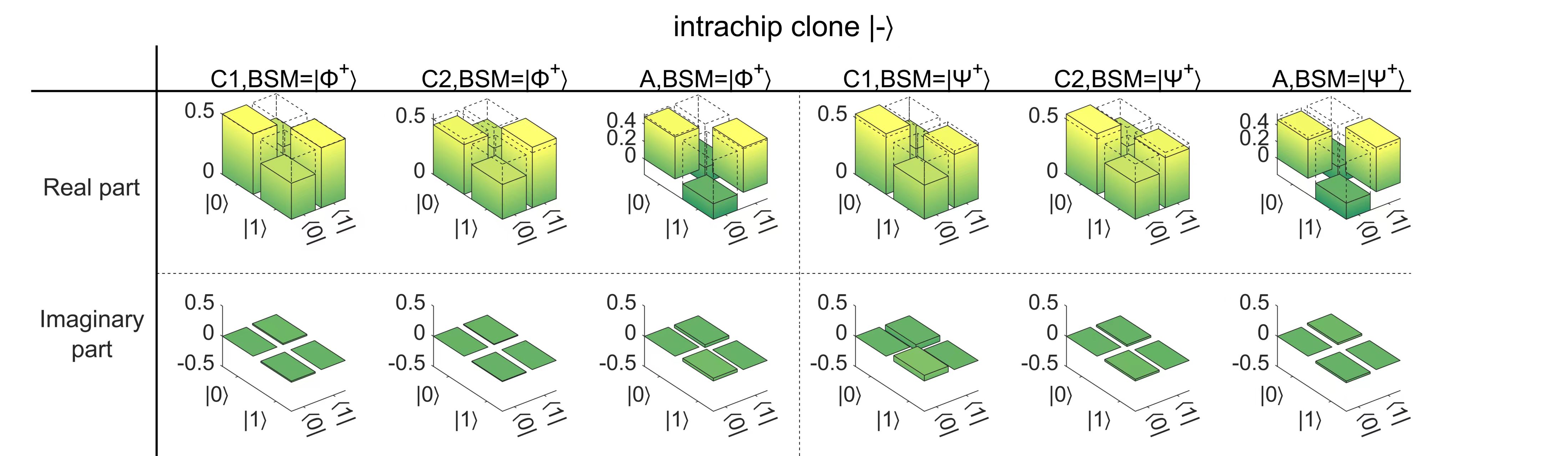}
    \caption{Reconstructed density matrices of photons $\mathrm{C}_1$, $\mathrm{C}_2$, and A for the cloned $|-\rangle$ state (intrachip configuration). The panels display the real and imaginary parts of the density matrices, reconstructed via MLE from parallel single-photon tomography data. The results are categorized by the BSM outcomes: $\ket{\Phi^+}$ and $\ket{\Psi^+}$. Within each category, the reconstructed states for the three individual photons are presented to verify the performance of the optimal symmetric telecloning process.}
    \label{intram}
\end{figure}

\begin{figure}[H]
    \centering
    \includegraphics[width=0.8\textwidth]{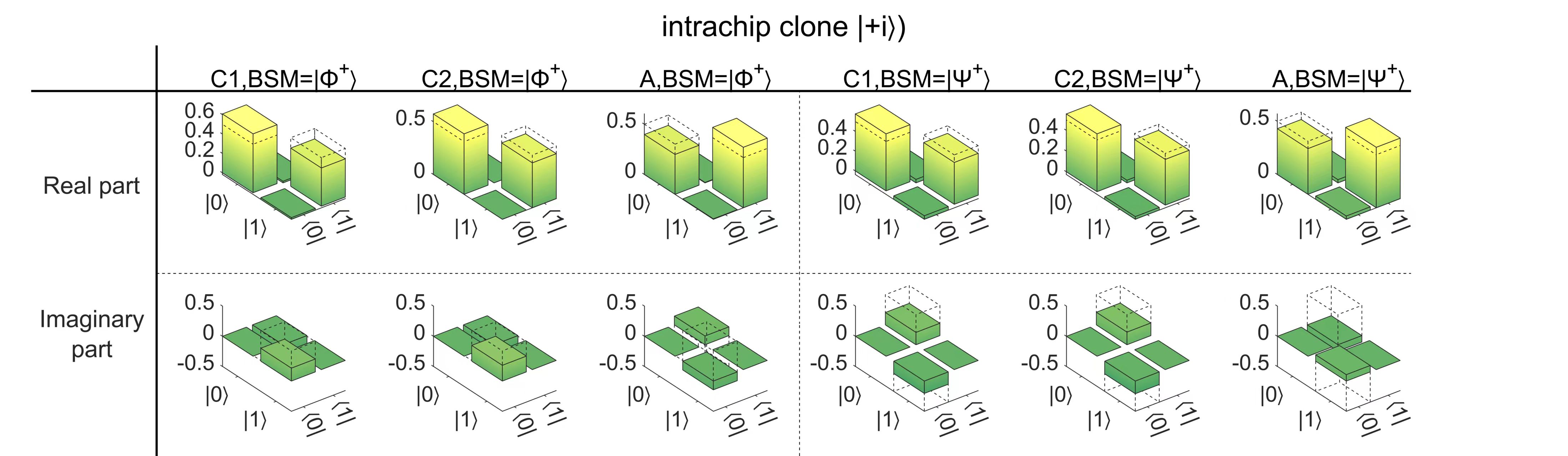}
    \caption{Reconstructed density matrices of photons $\mathrm{C}_1$, $\mathrm{C}_2$, and A for the cloned $|+i\rangle$ state (intrachip configuration). The panels display the real and imaginary parts of the density matrices, reconstructed via MLE from parallel single-photon tomography data. The results are categorized by the BSM outcomes: $\ket{\Phi^+}$ and $\ket{\Psi^+}$. Within each category, the reconstructed states for the three individual photons are presented to verify the performance of the optimal symmetric telecloning process.}
    \label{intrapi}
\end{figure}

\begin{figure}[H]
    \centering
    \includegraphics[width=0.8\textwidth]{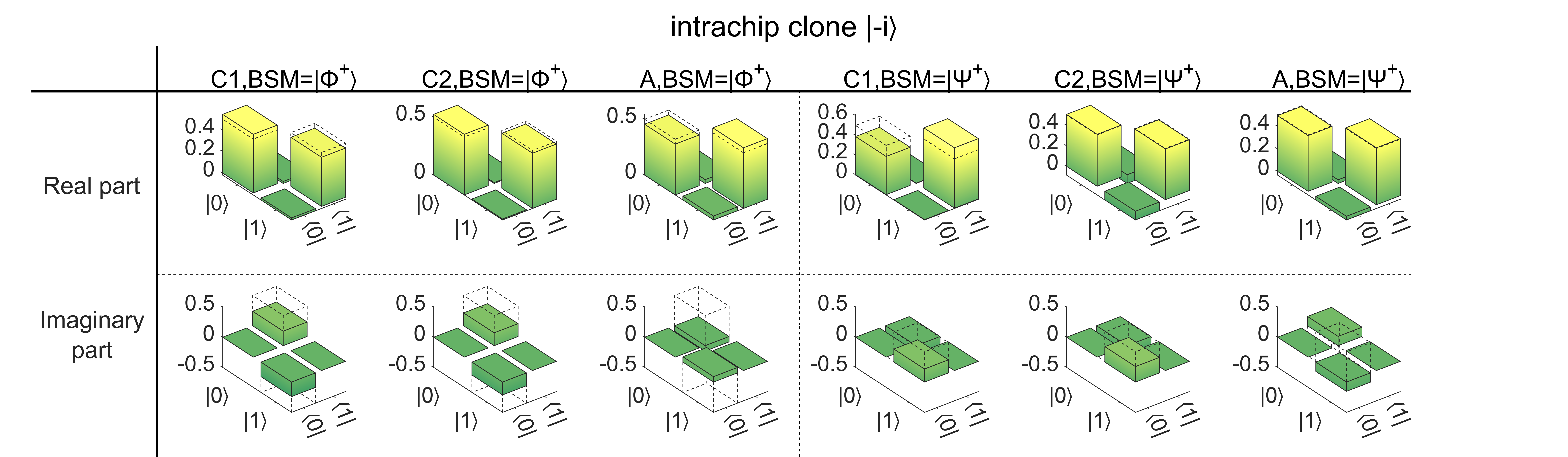}
    \caption{Reconstructed density matrices of photons $\mathrm{C}_1$, $\mathrm{C}_2$, and A for the cloned $|-i\rangle$ state (intrachip configuration). The panels display the real and imaginary parts of the density matrices, reconstructed via MLE from parallel single-photon tomography data. The results are categorized by the BSM outcomes: $\ket{\Phi^+}$ and $\ket{\Psi^+}$. Within each category, the reconstructed states for the three individual photons are presented to verify the performance of the optimal symmetric telecloning process.}
    \label{intrami}
\end{figure}

\begin{figure}[H]
    \centering
    \includegraphics[width=0.8\textwidth]{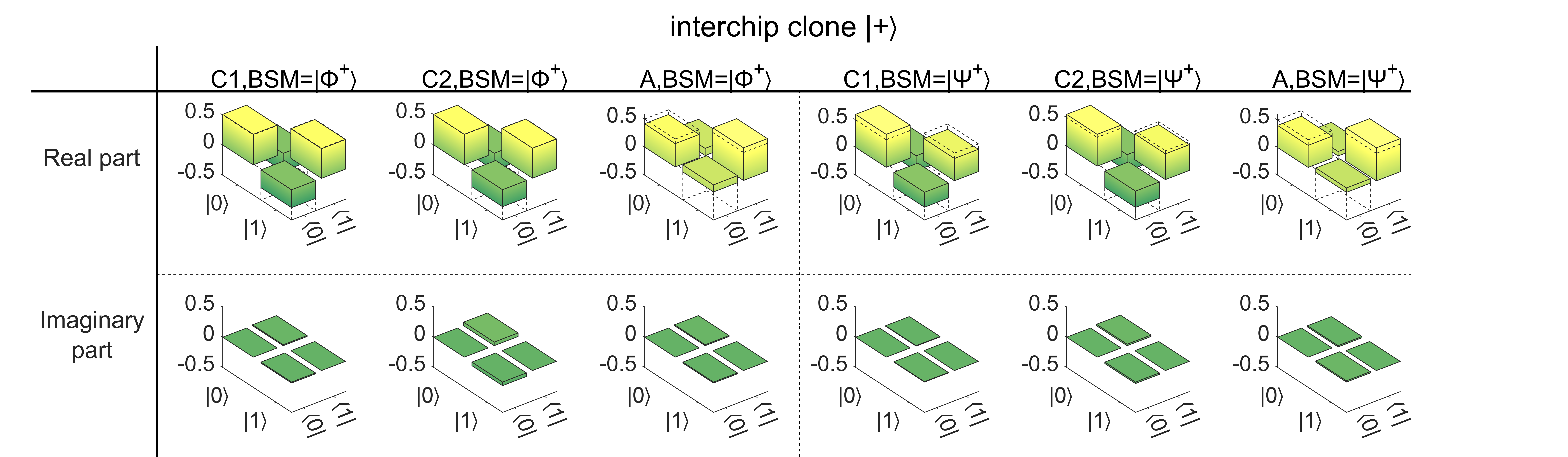}
    \caption{Reconstructed density matrices of photons $\mathrm{C}_1$, $\mathrm{C}_2$, and A for the cloned $|+\rangle$ state (interchip configuration). The panels display the real and imaginary parts of the density matrices, reconstructed via MLE from parallel single-photon tomography data. The results are categorized by the BSM outcomes: $\ket{\Phi^+}$ and $\ket{\Psi^+}$. Within each category, the reconstructed states for the three individual photons are presented to verify the performance of the optimal symmetric telecloning process.}
    \label{interp}
\end{figure}

To account for the on-chip multiport splitting introduced by the MMI devices, we treat the signal and idler outputs of sources S\textsubscript{2} and S\textsubscript{3} as being evenly divided into two paths. As illustrated in Fig. \ref{Chip2}, the outputs are routed as follows: the signal from S\textsubscript{2} is split to paths P\textsubscript{0} and A\textsubscript{0} while the idler from S\textsubscript{2} is split to C\textsubscript{1-1} and C\textsubscript{2-1}. Similarly, the signal from S\textsubscript{3} is split to P\textsubscript{1} and A\textsubscript{1} and the idler to C\textsubscript{1-0} and C\textsubscript{2-0}. 

Assuming ideal symmetric (50:50) MMIs, each photon is routed to one of the two available outputs with a probability of 1/2. Consequently, a specific signal-idler pair combination from the same source occurs with a probability of 1/4. This results in four distinct coincidence combinations per source. For S\textsubscript{2}, these are (P\textsubscript{0}, C\textsubscript{1-1}), (P\textsubscript{0}, C\textsubscript{2-1}), (A\textsubscript{0}, C\textsubscript{1-1}), (A\textsubscript{0}, C\textsubscript{2-1}). For S\textsubscript{3}, these are (P\textsubscript{1}, C\textsubscript{1-0}), (P\textsubscript{1}, C\textsubscript{2-0}), (A\textsubscript{1}, C\textsubscript{1-0}), (A\textsubscript{1}, C\textsubscript{2-0}). 

The single-count rate for any individual output path $m\in$\{P\textsubscript{0}, P\textsubscript{1}, A\textsubscript{0}, A\textsubscript{1}, C\textsubscript{1-0}, C\textsubscript{1-1}, C\textsubscript{2-0}, C\textsubscript{2-1}\} is given by:
\begin{equation} SC_{m} = DC_{m} +\frac{f[p\eta_m + p^2(2\eta_m-\eta_m^2)]}{2}\end{equation}
and the coincidence rate for a specific pair of output paths m,n originating from the same source becomes:
\begin{equation} CC_{mn} = ACC + \frac{pf\eta_m\eta_n}{4}\end{equation}
where the factor 1/4 accounts for the two independent 50:50 splits (signal and idler) that select the particular (m,n) combination.

As the MMI evenly distributes the photon flux, the measured single counts per branch are approximately half of the unsplit rate, and the coincidence rate for any given branch pair is reduced by a factor of four. Based on the single and coincidence counts listed in Table \ref{tab:combined_counts}, we extract the average generation probabilities $p_2=0.0383$ and $p_3=0.0436$ for S\textsubscript{2} and S\textsubscript{3}, respectively.

The calculated losses are summarized in Table \ref{tab:loss}. The average on-chip loss is $\sim$8.5 dB, while the inter-chip loss is $\sim$16.7 dB. For the on-chip (A, C\textsubscript{1}, C\textsubscript{2}) paths, the estimated loss of $\sim$8.1 dB includes the intrinsic source loss (1 cm waveguide, $\sim$1 dB), transmission waveguides (4mm, 2 dB/cm), seven MMIs ($\le 0.31\text{ dB}$ each), zero or one crossing ($\le 0.12\text{ dB}$), facet coupling ($\sim$70\% efficiency), WDM filters ($\sim$1.05 dB, see Table \ref{tab:loss_wdm}), three fiber adapters ($\sim$0.3 dB each), and the SSPD quantum efficiency ($\sim$90\%). For the inter-chip (P) paths, the estimated total loss of $\sim$15.72 dB further accounts for additional MMIs, crossings, PBRC ($\sim$0.43 dB), and PBRS ($\sim$0.2 dB) components. The minor discrepancies ($\sim$0.4--0.98 dB) between theoretical and measured values likely arise from variations in practical coupling efficiency compared to idealized test structures.

\subsubsection{Multi-fold coincidence rate}

Accounting for the post-selection efficiency, the theoretical four-fold coincidence rate ($CC_{4\text{-fold}}$) is determined as:

\begin{align}
\nonumber CC_{4\text{-}fold} &= \frac{1}{16}\,p_2 p_3 f\;(\eta_\mathrm{P_0}\eta_\mathrm{A_1}+\eta_\mathrm{P_1}\eta_\mathrm{A_0})
(\eta_\mathrm{C_{1\text{-}0}}\eta_\mathrm{C_{2\text{-}1}}+\eta_\mathrm{C_{1\text{-}1}}\eta_{C_{2\text{-}0}})\\[4pt]
&\quad +\frac{1}{4}\,p_2^2 f\,\eta_\mathrm{P_0}\eta_\mathrm{A_0}\eta_\mathrm{C_{1\text{-}1}}\eta_\mathrm{C_{2\text{-}1}}
+\frac{1}{4}\,p_3^2 f\,\eta_\mathrm{P_1}\eta_\mathrm{A_1}\eta_\mathrm{C_{1\text{-}0}}\eta_\mathrm{C_{2\text{-}0}}.
\end{align}

The calculated value of $CC_{4\text{-fold}} = 32.28\text{ Hz}$ is in good agreement with the experimental result of $36.39\text{ Hz}$. Furthermore, the six-fold coincidence rate is estimated as:$$CC_{6\text{-fold}} = CC_{4\text{-fold}} \cdot \frac{p_1\eta_X\eta_T}{2}$$This yields an estimated rate of $6.89/\text{hour}$, which is consistent with the experimentally observed rate of $10.47/\text{hour}$.

\clearpage

\onecolumngrid
\subsection{Supplementary Fig. S1-S13}

\clearpage

\subsection{Supplementary Table S1-S10}
\begin{table}[htbp]
\centering
\caption{Raw coincidence counts for the tomography of the cloned $|0\rangle$ state (intrachip configuration). The integration time is 69.24~h.}
\label{tab:cloning_counts1}
\begin{tabular}{cccccccc}
\toprule
BSM result & Photon & H & V & D & A & R & L \\ 
\midrule
\multirow{3}{*}{$\Phi^+$} & $\mathrm{C}_1$ & 62  & 186 & 136 & 130 & 147 & 125 \\
                          & $\mathrm{C}_2$ & 65  & 183 & 140 & 126 & 134 & 138 \\
                          & A  & 159 & 89  & 122 & 144 & 125 & 147 \\ 
\midrule
\multirow{3}{*}{$\Psi^+$} & $\mathrm{C}_1$ & 170 & 52  & 130 & 112 & 138 & 118 \\
                          & $\mathrm{C}_2$ & 159 & 63  & 132 & 110 & 137 & 119 \\
                          & A  & 92  & 130 & 100 & 142 & 118 & 138 \\ 
\bottomrule
\end{tabular}
\end{table}

\begin{table}[htbp]
\centering
\caption{Raw coincidence counts for the tomography of the cloned $|1\rangle$ state (intrachip configuration). The integration time is 97.5~h.}
\label{tab:cloning_counts_1}
\begin{tabular}{cccccccc}
\toprule
BSM result & Photon & H & V & D & A & R & L \\ 
\midrule
\multirow{3}{*}{$\Phi^+$} & $\mathrm{C}_1$ & 186 & 77 & 109 & 117 & 126 & 117 \\
                          & $\mathrm{C}_2$ & 196 & 67 & 118 & 108 & 148 & 95 \\
                          & A  & 103 & 160 & 109 & 117 & 105 & 138 \\ 
\midrule
\multirow{3}{*}{$\Psi^+$} & $\mathrm{C}_1$ & 67 & 178 & 126 & 134 & 128 & 135 \\
                          & $\mathrm{C}_2$ & 70 & 175 & 131 & 129 & 138 & 125 \\
                          & A  & 146 & 99 & 127 & 133 & 130 & 133 \\ 
\bottomrule
\end{tabular}
\end{table}

\begin{table}[htbp]
\centering
\caption{Raw coincidence counts for the tomography of the cloned $|+\rangle$ state (intrachip configuration). The integration time is 60.99~h.}
\label{tab:cloning_counts_plus}
\begin{tabular}{cccccccc}
\toprule
BSM result & Photon & H & V & D & A & R & L \\ 
\midrule
\multirow{3}{*}{$\Phi^+$} & $\mathrm{C}_1$ & 119 & 136 & 49 & 222 & 129 & 123 \\
                          & $\mathrm{C}_2$ & 112 & 143 & 62 & 209 & 127 & 125 \\
                          & A  & 139 & 116 & 169 & 102 & 129 & 123 \\ 
\midrule
\multirow{3}{*}{$\Psi^+$} & $\mathrm{C}_1$ & 111 & 158 & 50 & 216 & 110 & 115 \\
                          & $\mathrm{C}_2$ & 109 & 160 & 57 & 209 & 125 & 100 \\
                          & A  & 157 & 112 & 171 & 95 & 100 & 125 \\ 
\bottomrule
\end{tabular}
\end{table}

\begin{table}[htbp]
\centering
\caption{Raw coincidence counts for the tomography of the cloned $|-\rangle$ state (intrachip configuration). The integration time is 55.23~h.}
\label{tab:cloning_counts_minus}
\begin{tabular}{cccccccc}
\toprule
BSM result & Photon & H & V & D & A & R & L \\ 
\midrule
\multirow{3}{*}{$\Phi^+$} & $\mathrm{C}_1$ & 97 & 96 & 151 & 38 & 92 & 100 \\
                          & $\mathrm{C}_2$ & 87 & 106 & 154 & 35 & 94 & 98 \\
                          & A  & 93 & 100 & 61 & 128 & 105 & 87 \\ 
\midrule
\multirow{3}{*}{$\Psi^+$} & $\mathrm{C}_1$ & 101 & 89 & 159 & 40 & 111 & 76 \\
                          & $\mathrm{C}_2$ & 104 & 86 & 165 & 34 & 99 & 88 \\
                          & A  & 88 & 102 & 61 & 138 & 88 & 99 \\ 
\bottomrule
\end{tabular}
\end{table}

\begin{table}[htbp]
\centering
\caption{Raw coincidence counts for the tomography of the cloned $|+i\rangle$ state (intrachip configuration). The integration time is 36~h.}
\label{tab:cloning_counts_plus_i}
\begin{tabular}{cccccccc}
\toprule
BSM result & Photon & H & V & D & A & R & L \\ 
\midrule
\multirow{3}{*}{$\Phi^+$} & $\mathrm{C}_1$ & 148 & 96 & 117 & 125 & 197 & 79 \\
                          & $\mathrm{C}_2$ & 140 & 104 & 121 & 121 & 210 & 66 \\
                          & A  & 97 & 147 & 122 & 120 & 98 & 178 \\ 
\midrule
\multirow{3}{*}{$\Psi^+$} & $\mathrm{C}_1$ & 158 & 116 & 99 & 114 & 60 & 143 \\
                          & $\mathrm{C}_2$ & 153 & 121 & 101 & 112 & 59 & 144 \\
                          & A  & 118 & 156 & 113 & 100 & 123 & 80 \\ 
\bottomrule
\end{tabular}
\end{table}

\begin{table}[htbp]
\centering
\caption{Raw coincidence counts for the tomography of the cloned $|-i\rangle$ state (intrachip configuration). The integration time is 41.49~h.}
\label{tab:cloning_counts_minus_i}
\begin{tabular}{cccccccc}
\toprule
BSM result & Photon & H & V & D & A & R & L \\ 
\midrule
\multirow{3}{*}{$\Phi^+$} & $\mathrm{C}_1$ & 105 & 88 & 91 & 98 & 38 & 108 \\
                          & $\mathrm{C}_2$ & 101 & 92 & 96 & 93 & 42 & 104 \\
                          & A  & 88 & 105 & 101 & 88 & 86 & 60 \\ 
\midrule
\multirow{3}{*}{$\Psi^+$} & $\mathrm{C}_1$ & 69 & 109 & 85 & 86 & 126 & 50 \\
                          & $\mathrm{C}_2$ & 90 & 88 & 71 & 100 & 133 & 43 \\
                          & A  & 88 & 90 & 80 & 91 & 61 & 115 \\ 
\bottomrule
\end{tabular}
\end{table}

\begin{table}[htbp]
\centering
\caption{Raw coincidence counts for the tomography of the cloned $|+\rangle$ state (interchip configuration).The integration time is 162.5001~h.}
\label{tab:cloning_counts_interchip_plus}
\begin{tabular}{cccccccc}
\toprule
BSM result & Photon & H & V & D & A & R & L \\ 
\midrule
\multirow{3}{*}{$\Phi^+$} & $\mathrm{C}_1$ & 140 & 136 & 54 & 211 & 123 & 131 \\
                          & $\mathrm{C}_2$ & 139 & 137 & 58 & 207 & 112 & 142 \\
                          & A  & 115 & 161 & 166 & 99 & 123 & 131 \\ 
\midrule
\multirow{3}{*}{$\Psi^+$} & $\mathrm{C}_1$ & 188 & 127 & 69 & 231 & 148 & 144 \\
                          & $\mathrm{C}_2$ & 174 & 141 & 65 & 235 & 139 & 153 \\
                          & A  & 126 & 189 & 174 & 126 & 152 & 140 \\ 
\bottomrule
\end{tabular}
\end{table}

\begin{table}[htbp]
\centering
\caption{Measured singles and coincidence count rates for different channels and pairs.}
\label{tab:combined_counts}
\small 
\begin{tabular}{lcccccccc}
\toprule
\textbf{Singles} & P\textsubscript{0} & P\textsubscript{1} & A\textsubscript{0} & A\textsubscript{1} & C\textsubscript{1-1} & C\textsubscript{1-0} & C\textsubscript{2-1} & C\textsubscript{2-0} \\ 
\midrule
Rate (Hz) & $2.55 \times 10^5$ & $1.89 \times 10^5$ & $1.09 \times 10^6$ & $1.43 \times 10^6$ & $1.29 \times 10^6$ & $1.70 \times 10^6$ & $1.35 \times 10^6$ & $1.81 \times 10^6$ \\
\midrule
\midrule
\textbf{Coinc.} & P\textsubscript{0}-C\textsubscript{1-1} & P\textsubscript{0}-C\textsubscript{2-1} & A\textsubscript{0}-C\textsubscript{1-1} & A\textsubscript{0}-C\textsubscript{2-1} & P\textsubscript{1}-C\textsubscript{1-0} & P\textsubscript{1}-C\textsubscript{2-0} & A\textsubscript{1}-C\textsubscript{1-0} & A\textsubscript{1}-C\textsubscript{2-0} \\ 
\midrule
Rate (Hz) & $1.63 \times 10^4$ & $1.84 \times 10^4$ & $7.41 \times 10^4$ & $7.83 \times 10^4$ & $1.38 \times 10^4$ & $1.46 \times 10^4$ & $1.09 \times 10^5$ & $1.43 \times 10^5$ \\
\bottomrule
\end{tabular}
\end{table}

\begin{table}[htbp]
\centering
\caption{Calculated path losses.}
\label{tab:loss}
\begin{tabular}{ccccccccc}
\toprule
\textbf{Channel} & $\eta_\mathrm{P_0}$ & $\eta_\mathrm{P_1}$ & $\eta_\mathrm{A_0}$ & $\eta_\mathrm{A_1}$ 
& $\eta_\mathrm{C_{1-1}}$ & $\eta_\mathrm{C_{1-0}}$ & $\eta_\mathrm{C_{2-1}}$ & $\eta_\mathrm{C_{2-0}}$ \\
\midrule
Loss (dB) & 15.80 & 17.69 & 9.24 & 8.97 
& 8.76 & 8.15 & 8.24 & 7.89 \\
\bottomrule
\end{tabular}
\end{table}

\begin{table}[htbp]
\centering
\caption{WDM loss}
\label{tab:loss_wdm}
\begin{tabular}{cccccccc}
\toprule
$\eta_\mathrm{P_0}^{\mathrm{WDM}}$ &
$\eta_\mathrm{P_1}^{\mathrm{WDM}}$ &
$\eta_\mathrm{A_0}^{\mathrm{WDM}}$ &
$\eta_\mathrm{A_1}^{\mathrm{WDM}}$ &
$\eta_\mathrm{C_{1-1}}^{\mathrm{WDM}}$ &
$\eta_\mathrm{C_{1-0}}^{\mathrm{WDM}}$ &
$\eta_\mathrm{C_{2-1}}^{\mathrm{WDM}}$ &
$\eta_\mathrm{C_{2-0}}^{\mathrm{WDM}}$ \\
\midrule
1.17\,dB & 1.06\,dB & 1.28\,dB & 0.81\,dB &
1.06\,dB & 0.98\,dB & 1.00\,dB & 1.03\,dB \\
\bottomrule
\end{tabular}
\end{table}

\twocolumngrid
\clearpage
\bibliography{teleclone.bib}

\end{document}